\documentclass[twocolumn,showpacs,amsmath,amssymb,aps,prd,floats]{revtex4-1}
\usepackage{graphicx}% Include figure files
\usepackage[switch,pagewise,running]{lineno}
\usepackage{dcolumn}% Align table columns on decimal point
\usepackage{bm}% bold math
\usepackage{epsfig}
\usepackage{xspace}
\usepackage{hyperref}
\hypersetup{colorlinks,linkcolor=red,citecolor=blue,urlcolor=blue}  
\usepackage{color,xcolor}
\usepackage{verbatim}
\usepackage{float}
\usepackage{slashed}
\usepackage{rotating}
\usepackage[normalem]{ulem} 
\usepackage[caption=false]{subfig}
\usepackage{verbatim}

\begin{document}

\title{Revisiting the proton synchrotron radiation in blazar jets: Possible contributions from X-ray to $\gamma$-ray bands}
\author{Rui Xue$^{1}$}
\author{Shao-Teng Huang$^{1}$}
\author{Hu-Bing Xiao$^{2}$}\email{E-mail: hubing.xiao@shnu.edu.cn}
\author{Ze-Rui Wang$^{3}$}\email{zerui\_wang62@163.com}

\affiliation{$^1$Department of Physics, Zhejiang Normal University, Jinhua 321004, China\\
$^2$Shanghai Key Lab for Astrophysics, Shanghai Normal University, Shanghai 200234, People's Republic of China\\
$^3$College of Physics and Electronic Engineering, Qilu Normal University, Jinan 250200, China}

\begin{abstract}
The proton synchrotron radiation is considered as the origin of high-energy emission of blazars at times. However, extreme physical parameters are often required. In this work, we propose an analytical method to study the parameter space when applying the proton synchrotron radiation to fit the keV, GeV, and very-high-energy emission of blazar jets. We find that proton synchrotron radiation can fit the high-energy hump when it peaks beyond tens GeV without violating basic observations and theories. For the high-energy hump peaked around GeV band, extreme parameters, such as a super-Eddington jet power and a very strong magnetic field, are required. For the high-energy hump peaked around keV band, if an acceptable parameter space can be found depends on the object's keV luminosity.
\end{abstract}

\maketitle

\section{Introduction}
The past few decades have been a time of tremendous progress in search for high-energy $\gamma$-ray extragalactic sources. Numerous objects have been discovered due to the development of the latest generation of observational equipment, especially the $Fermi$ $Gamma$-$ray$ $Space$ $Telescope$ \cite{2020ApJS..247...33A}. The major population of the extragalactic $\gamma$-ray bright objects is blazars, a subclass of active galactic nuclei (AGNs) with relativistic jets pointed along the observers' line of sight \cite{1995PASP..107..803U,2019Galax...7...23R}. The multi-wavelength spectral energy distribution (SED) of blazars is dominated by jets' non-thermal emission, which usually exhibit a two-hump structure. Conventionally, the low-energy hump, from radio to UV/X-ray, is thought to be originated from the synchrotron radiation of accelerated relativistic electrons, while the modeling of high-energy hump, from X-ray to $\gamma$-ray, leads to bifurcated leptonic models and hadronic models. In leptonic models, the high-energy hump is explained by inverse Compton (IC) radiation from relativistic electrons that up-scatter soft photons emitted by the same electrons population (synchrotron-self Compton, SSC; \cite{1985ApJ...298..114M,1992ApJ...397L...5M,1992A&A...256L..27D}), or soft photons originated from external photon fields (external Compton, EC; \cite{1993ApJ...416..458D}). In hadronic models, the high-energy hump is interpreted by many processes \cite{2000NewA....5..377A,2001APh....15..121M, 2002MNRAS.332..215A, 2013ApJ...768...54B, 2015MNRAS.448..910C, 2017MNRAS.464.2213P, 2019ApJ...881...46R, 2020ApJ...889..149D, 2022A&A...658L...6D, 2022PhRvD.106j3021X}, among which the proton synchrotron radiation is the commonly considered one.

When applying proton synchrotron radiation to explain the high-energy hump of blazars, some extreme physical parameters are often introduced. Among which the most widely discussed is the total jet power. In general, the Eddington luminosity of the central supermassive black hole (SMBH) is treated as the upper limit of the total jet power \cite{2015MNRAS.453.3213S, 2015MNRAS.450L..21Z, 2019ApJ...871...81X}. Some analytical and modeling studies suggest that a highly super-Eddington jet power has to be introduced because of the large particle mass of protons and the relative low radiation efficiency \cite{2013ApJ...768...54B, 2015MNRAS.450L..21Z}. However, some works put forward different opinions. Through analytical analysis, \cite{2016ApJ...825L..11P} suggest that the proton synchrotron radiation model when explaining the very-high-energy (VHE; $\geqslant 0.1~\rm TeV$) emission can have a sub-Eddington jet power, while a super-Eddington jet power is still needed when explaining the powerful GeV emission. In the numerical fitting of VHE spectrum, \cite{2015MNRAS.448..910C} also find solutions with the sub-Eddington jet power by searching the parameter space in a wide range. Also, due to the relative low radiation efficiency of proton synchrotron emission, a strong magnetic field ($\gg 10~\rm G$) is usually required in the theoretical modeling \cite{2000NewA....5..377A, 2013ApJ...768...54B, 2016ApJ...825L..11P}. However, radio observations and polarization measurements indicate that the sub-pc scale jet usually has a weak magnetic field ($< 10~\rm G$; e.g., \cite{2009MNRAS.400...26O, 2012A&A...545A.113P, 2016A&A...590A..48K, 2017A&A...597A..80H, 2021A&A...651A..74K, 2022MNRAS.510..815K, 2022MNRAS.509L..21L, 2023ApJ...942L..10M, 2023A&A...669A..32P}), which may not meet the requirements of the theoretical modeling. 

In this work, we develop an analytical method to revisit the proton synchrotron radiation in blazar jets. In addition to the widely discussed contribution at GeV and VHE bands, we also study if the proton synchrotron radiation can contribute to X-ray band. With the proposed analytical method, we will show if the proton synchrotron radiation can explain the emission of high-energy hump without introducing extreme physical parameters (e.g., a super-Eddington luminosity or a strong magnetic field described before). In section~\ref{method}, we present the analytical method to find the parameter space. In section~\ref{app}, we employ the analytical method to some well-known blazars, studying if a reasonable parameter space can be obtained. If a reasonable parameter space is obtained, we also fit the multi-wavelength SED with the one-zone proton synchrotron model. A summary is given in section~\ref{sum}. Throughout the paper, the cosmological parameters $H_{0}=69.6\ \rm km\ s^{-1}Mpc^{-1}$, $\Omega_{0}=0.29$, and $\Omega_{\Lambda}$= 0.71 \cite{2014ApJ...794..135B} are adopted in this work.

\section{Method}\label{method}
In the modeling of blazars' emission, it is widely assumed that all the non-thermal radiation of the jet comes from a compact spherical emitting region. It is composed of a plasma of charged particles in a uniformly magnetic field $B$ with radius $R$ and moving with bulk Lorentz factor $\Gamma=(1-\beta_{\Gamma}^2)^{-1/2}$, where $\beta_{\Gamma}c$ is the jet speed, along the jet, at a viewing angle $\theta_{\rm obs}$ with respect to observers' line of sight. Due to the beaming effect, the observed radiation is strongly boosted by a relativistic Doppler factor $\delta_{\rm D}=[\Gamma(1-\beta_{\Gamma} \rm cos\theta_{\rm obs})]^{-1}$. In this work, by assuming $\theta_{\rm obs} \lesssim 1/\Gamma$ for blazars, we have $\delta_{\rm D} \approx \Gamma$. 
In the following, the analytical study of constraining the parameter space is given under the framework of this conventional one-zone model. In this section, the parameters with superscript ``obs'' are measured in observers' frame, those with superscript ``AGN'' are measured in the AGN frame, whereas the parameters without the superscript are measured in the comoving frame, unless specified otherwise.

In the analytical calculation, there are three constraints on the parameter space, the first of which is that the total jet power $L_{\rm jet}^{\rm AGN}$ does not exceed the Eddington luminosity $L_{\rm Edd}$ of SMBH,
\begin{equation}\label{Lj}
L_{\rm Edd}\geqslant L_{\rm jet}^{\rm AGN} = L_{\rm p, cold}^{\rm AGN}+L_{\rm e, inj}^{\rm AGN}+L_{\rm B}^{\rm AGN}+L_{\rm p, inj}^{\rm AGN},
\end{equation}
where the four terms on the right-hand side represent the kinetic power in cold protons, the injection power of relativistic electrons, the power carried in magnetic field, and the injection power of relativistic protons. In this work, it is assumed that the high-energy hump is mainly from the proton synchrotron radiation, therefore the total jet power is dominated by the power of injected relativistic protons $L_{\rm p, inj}^{\rm AGN}$ and magnetic field $L_{\rm B}^{\rm AGN}$. More specifically, 
\begin{equation}\label{LB}
L_{\rm B}^{\rm AGN}=\pi R^2\Gamma^2cU_{\rm B},
\end{equation}
where $c$ is speed of light, $U_{\rm B}=B^2/8\pi$ is the magnetic field energy density, and $L_{\rm p, inj}^{\rm AGN}=L_{\rm p, inj}\Gamma^2$, where $L_{\rm p, inj}$ is the injection power of relativistic protons in the comoving frame. If assuming that relativistic protons are injected with a power-law energy distribution and the spectral index is 2 as predicted by the diffuse shock acceleration \cite{2007Ap&SS.309..119R}, we have 
\begin{equation}\label{Lp}
L_{\rm p, inj} = \frac{{\nu L_\nu}_{\rm peak}^{\rm obs}\Gamma^2 \rm{ln}\gamma_{\rm p, max}}{f_{\rm p, syn}\delta_{\rm D}^4},
\end{equation}
where ${\nu L_\nu}_{\rm peak}^{\rm obs}$ is the observed peak luminosity of the high-energy hump, $\gamma_{\rm p, max}$ can be deduced from the monochromatic approximation
\begin{equation}\label{gp}
\gamma_{\rm p, max} = (\frac{E_{\rm peak}^{\rm obs}(1+z)}{1.53\times10^3hB\delta_{\rm D}})^{1/2},
\end{equation}
where $h$ is the Planck constant, $z$ is the redshift and $E_{\rm peak}^{\rm obs}$ is the observed peak energy of the high-energy hump, 
and 
\begin{equation}\label{fp}
f_{\rm p, syn} = \rm min\{ \frac{t_{\rm dyn}}{t_{\rm p, syn}}, 1 \} = min\{ \frac{\sigma_{\rm T}\textit{B}^2\textit{R}\gamma_{\rm p, max}}{6\pi m_{\rm e} \textit{c}^2(\frac{\textit{m}_{\rm p}}{\textit{m}_{\rm e}})^3}, 1 \}
\end{equation}
is the cooling efficiency of proton synchrotron radiation, where $t_{\rm dyn}$ is the dynamical timescale, $t_{\rm p, syn}$ is the cooling timescale of proton synchrotron radiation, $\sigma_{\rm T}$ is the Thomson scattering cross section, $m_{\rm e}$ is the rest mass of an electron, and $m_{\rm p}$ is the rest mass of a proton. Substituting Eqs. (\ref{LB}), (\ref{Lp}), (\ref{gp}) and (\ref{fp}) into Eq.~(\ref{Lj}), it can be found that there are three parameters, which are $B$, $R$, and $\delta_{\rm D}$, respectively. By fixing the blob radius $R$ which can be inferred from the observed minimum variability timescale, we can find the parameter space of $B$ and $\delta_{\rm D}$ satisfying the constraint $L_{\rm Edd}\geqslant L_{\rm jet}^{\rm AGN}$.

In the above calculation, $\gamma_{\rm p, max}$ is deduced from the observed peak energy of the high-energy hump. However it is necessary to check if protons can be accelerated to this energy, which is the second constraint limiting the parameter space. Here, we estimate the maximum proton energy from the Hillas condition \cite{1984ARA&A..22..425H}
\begin{equation}\label{hillas}
E_{\rm p, max}^{\rm obs}\simeq 10^{21}\frac{B}{1~\rm G}\frac{R}{1~\rm pc}\frac{\delta_{\rm D}}{1+z}~\rm eV.
\end{equation}
Similarly, fixing the value of $R$, we can find the parameter space of $B$ and $\delta_{\rm D}$  when protons can be accelerated to the required energy.

Ensuring that the emitting region is transparent to high-energy emission is the third condition limiting the parameter space, which is important for the VHE emission. Using the $\delta$-approximation, the internal $\gamma \gamma$ opacity can be estimated as \cite{2009herb.book.....D}
\begin{equation}\label{tau}
\tau_{\gamma \gamma}\approx \frac{\sigma_{\gamma \gamma}{\nu L_\nu}_{\rm low}^{\rm obs}(E_{\rm low}^{\rm obs})}{4\pi RcE_{\rm low}^{\rm obs}\delta_{\rm D}^3}<1,
\end{equation}
where $\sigma_{\gamma\gamma} \approx 1.68 \times 10^{-25}~\rm cm^2$ is the cross section for $\gamma\gamma$ annihilation, $E_{\rm low}^{\rm obs}=\frac{2m_{\rm e}c^2\delta_{\rm D}^2}{(1+z)^2E_{\rm peak, obs}}$ represents energy of target photons interacted with the photons at the peak of the high-energy hump, and ${\nu L_\nu}_{\rm low}^{\rm obs}(E_{\rm low}^{\rm obs})$ is the corresponding luminosity. For a specific blazar, we can get the values of $E_{\rm low}^{\rm obs}$ and ${\nu L_\nu}_{\rm low}^{\rm obs}(E_{\rm low}^{\rm obs})$ from its SED. Then, fixing the value of $R$, the range of $\delta_{\rm D}$ that makes the emitting region transparent can be found. 

%As described above, our analytical method consists of three constraints. When the radius of the emitting region is given, we can find the parameter space of the magnetic field and the Doppler factor.

\section{Application}\label{app}
In this section, we apply the analytical method, i.e., Eqs.~(\ref{Lj}), (\ref{hillas}), and (\ref{tau}), to several well-known blazars, showing the ratio of $L_{\rm jet}/L_{\rm Edd}$ in the $\delta_{\rm D}$--$B$ diagram for different radius of emitting region. As indicated by observations, if the parameter space can be found when $B\lesssim10~\rm G$ \cite{2009MNRAS.400...26O, 2012A&A...545A.113P, 2016A&A...590A..48K, 2017A&A...597A..80H, 2021A&A...651A..74K, 2022MNRAS.510..815K} and $\delta_{\rm D}\lesssim30$ \cite{2009A&A...494..527H} (hereafter referred to as the observational constraints), we also fit their SEDs with the obtained reasonable values of $B$, $\delta_{\rm D}$ and $R$. A detailed numerical model description can be found in our previous work~\cite{2022PhRvD.106j3021X}.  To better fit the low-energy hump, relativistic electrons are assumed to be injected with a broken power-law energy distribution. For a self-consistent comparison with the analytical result, relativistic protons are assumed to be injected in the form of a power-law energy distribution with a spectral index of 2. Note that, hadronic interaction processes, including $p\gamma$ and $pp$, are not considered in the modeling, since many works argue that they are normally inefficient in the framework of one-zone model \cite{2009ApJ...704...38S, 2022A&A...659A.184L}. 

\begin{figure*}
\centering
\subfloat{
\includegraphics[width=1\columnwidth]{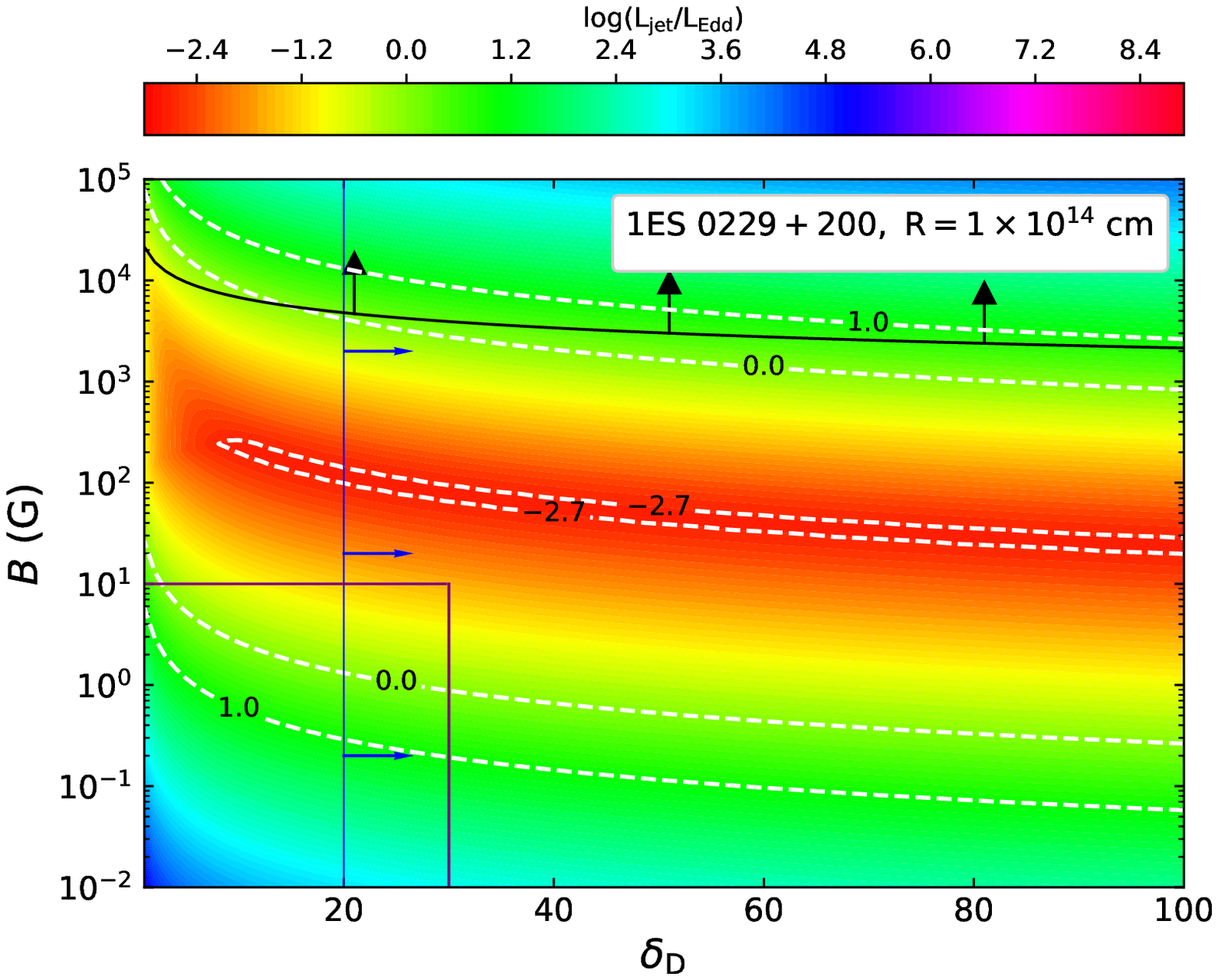}
}\hspace{-5mm}
\vspace{-1mm}
\quad
\subfloat{
\includegraphics[width=1\columnwidth]{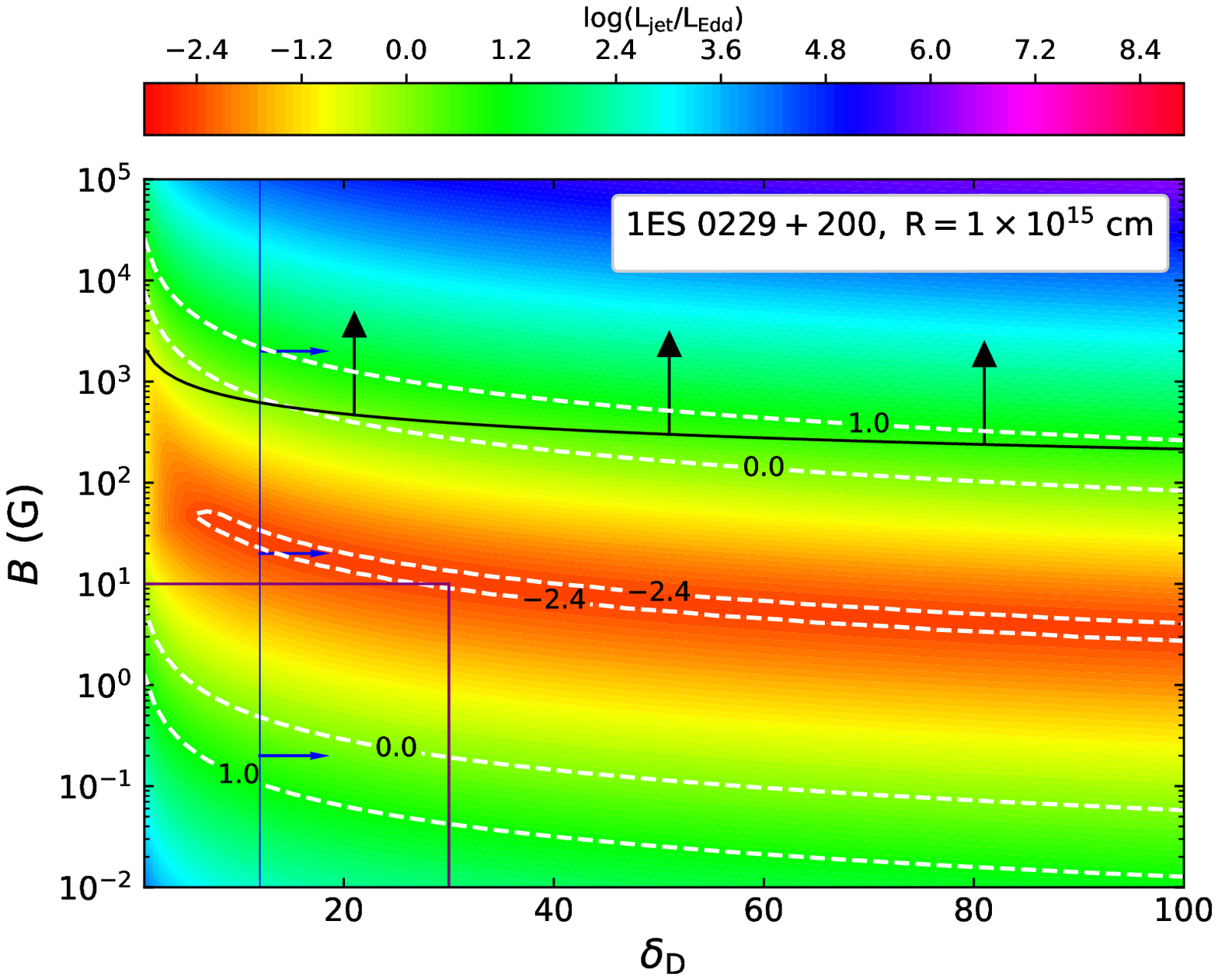}
}\hspace{-5mm}
\vspace{-1mm}
\quad
\subfloat{
\includegraphics[width=0.99\columnwidth]{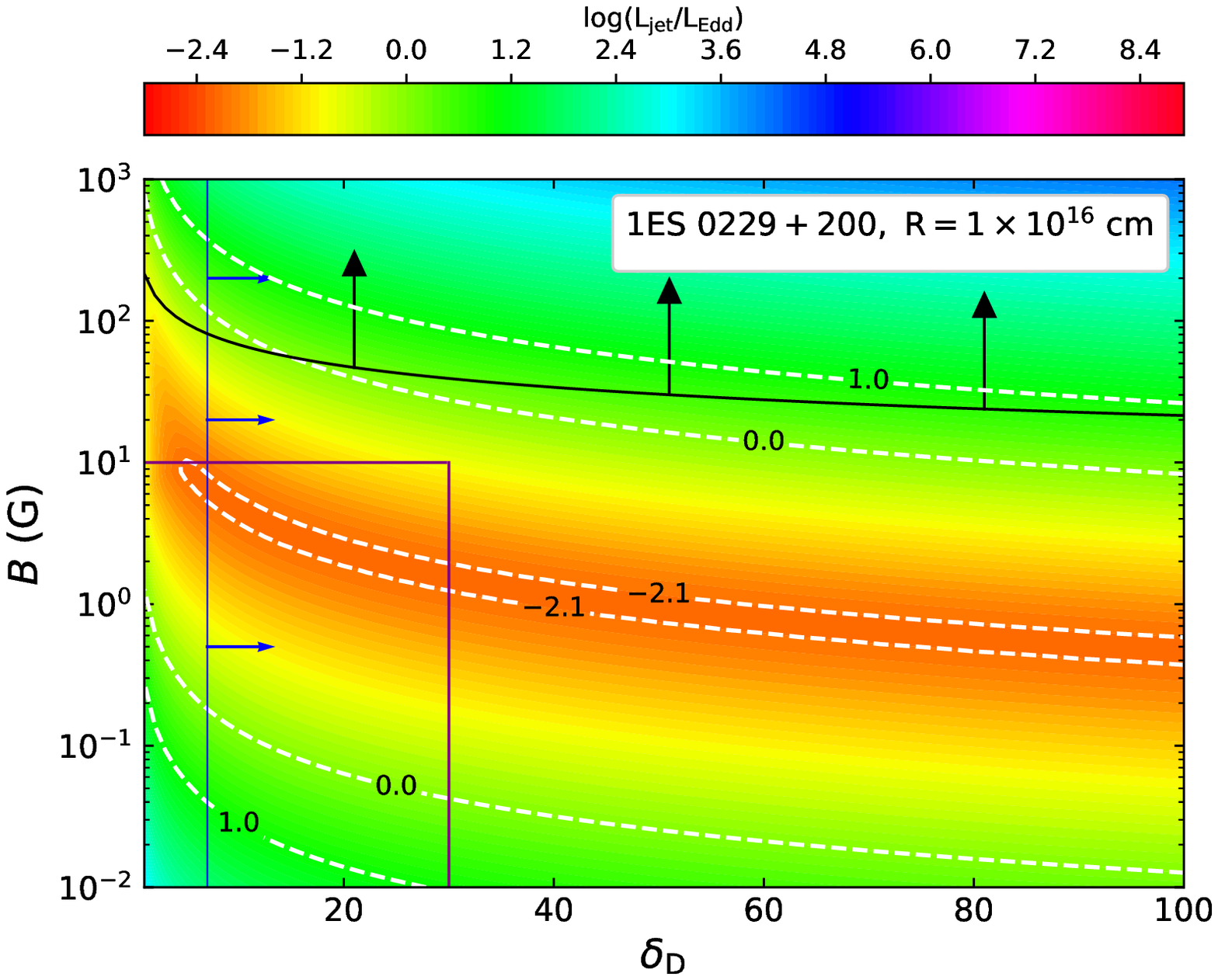}
}\hspace{-5mm}
\vspace{-1mm}
\quad
\subfloat{
\includegraphics[width=0.99\columnwidth]{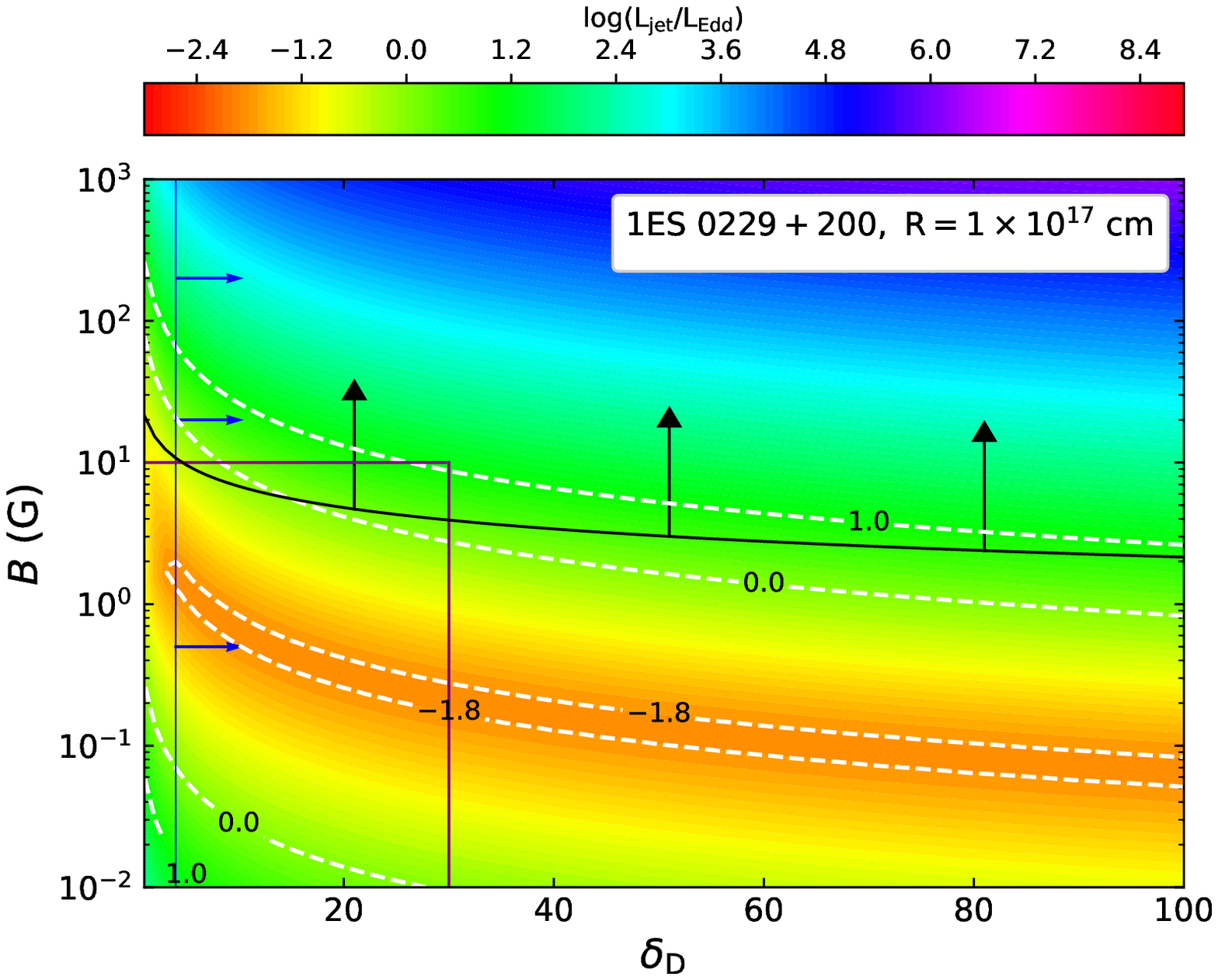}
}\hspace{-5mm}
\vspace{-1mm}
\quad
\subfloat{
\includegraphics[width=0.99\columnwidth]{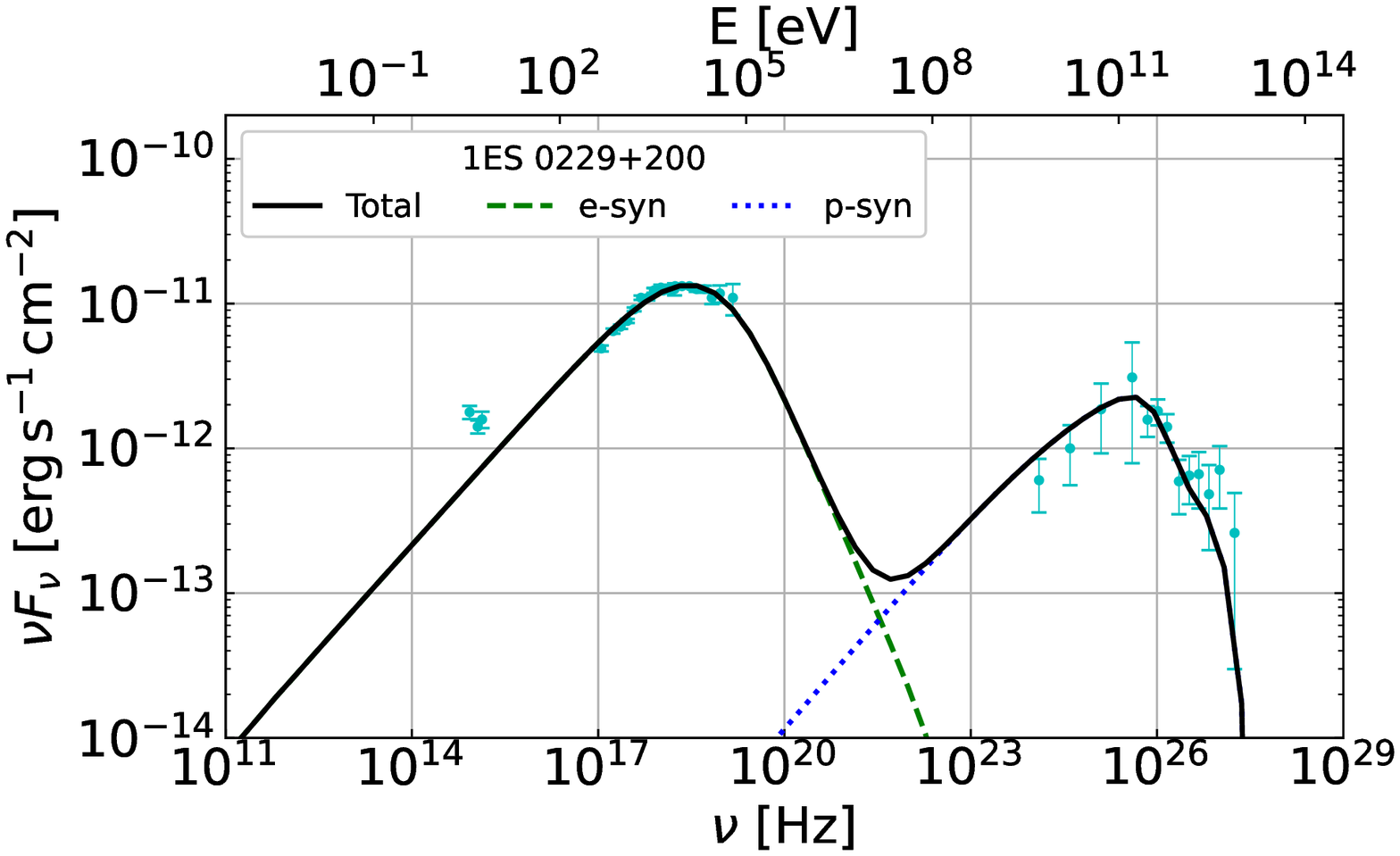}
}\hspace{-5mm}
\vspace{-1mm}
\quad
\subfloat{
\includegraphics[width=0.99\columnwidth]{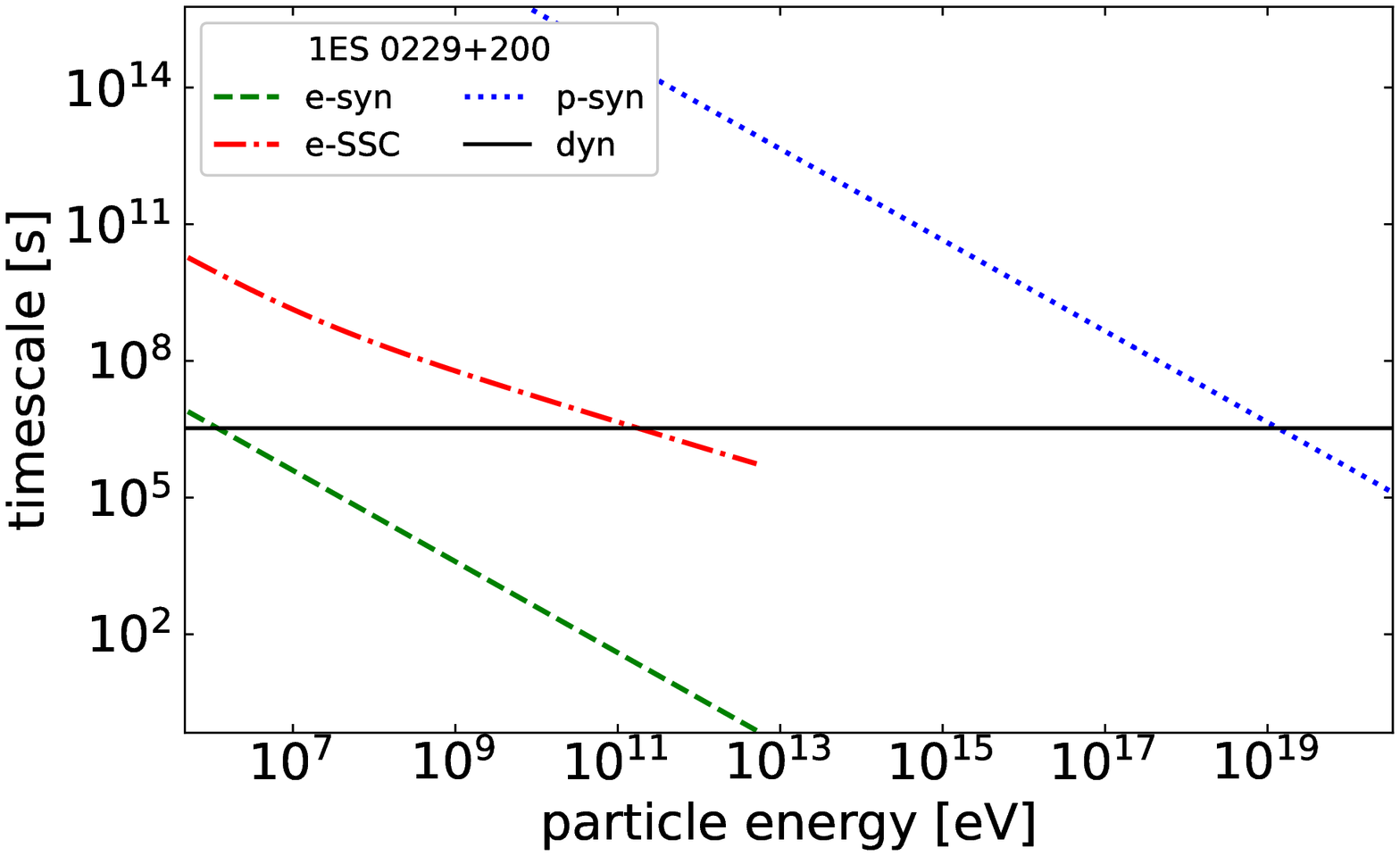}
}
\caption{Upper four panels: The ratio of $L_{\rm jet}/L_{\rm Edd}$ in the $\delta_{\rm D}$-$B$ diagram for 1ES 0229+200 peaking at 7.3~TeV, when setting $R=1\times 10^{14},~1\times 10^{15},~1\times 10^{16},~1\times 10^{17}~\rm cm$, respectively. The black curves with arrows represent the parameter space that satisfied the Hillas condition. The vertical blue curves with arrows show the lower limit of $\delta_{\rm D}$ that allows the escape of maximum energy $\gamma$-ray photons. The vertical and horizontal purple curves show the space that $B\lesssim10~\rm G$ and $\delta_{\rm D}\lesssim30$. The white dashed contours denote specific values of log($L_{\rm jet}/L_{\rm Edd}$) associated with the color bar. Lower two panels: The fitting result of the SED of 1ES 0229+200 with the conventional one-zone model (left panel) and the corresponding timescales of various cooling processes for relativistic electrons and protons as a function of particle energy in the comoving frame (right panel). In the lower left panel, the cyan points from UV to GeV bands and in the VHE band are quasi-simultaneous data taken from \cite{2018MNRAS.477.4257C}  and archival data taken from \cite{2014ApJ...782...13A}, respectively. The dashed green and blue dotted curves represent the relativistic electrons and protons synchrotron emissions, respectively. The solid black curve is the total emission from the emitting region. In the lower right panel, the green dashed, red dash-dotted, blue dotted, and black solid curves represent the relativistic electrons synchrotron, SSC, protons synchrotron, and dynamical timescales, respectively. The parameters are the same as those shown in Table~\ref{parameters}. \label{0229}}
\end{figure*}

\begin{figure*}
\centering
\subfloat{
\includegraphics[width=1\columnwidth]{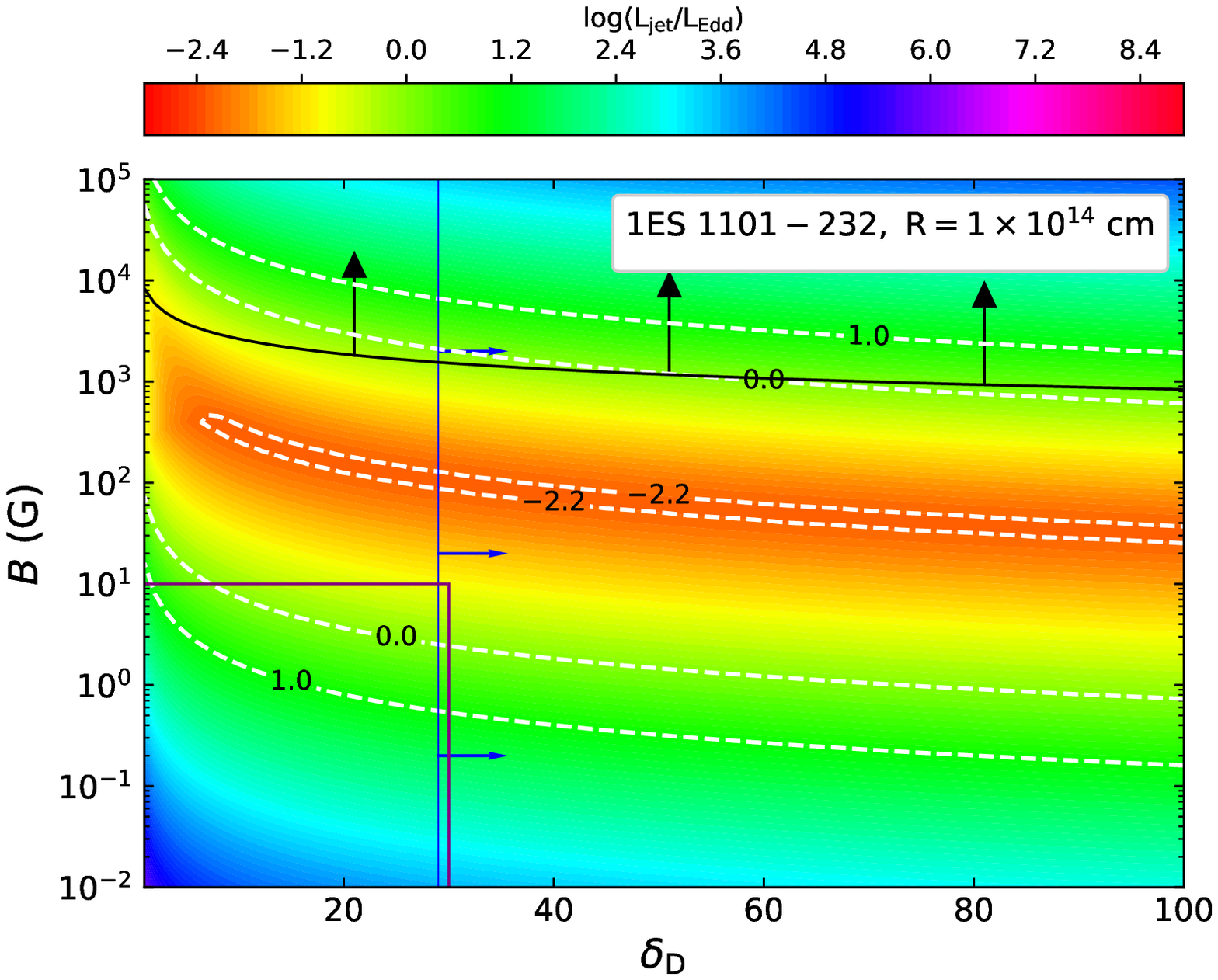}
}\hspace{-5mm}
\vspace{-1mm}
\quad
\subfloat{
\includegraphics[width=1\columnwidth]{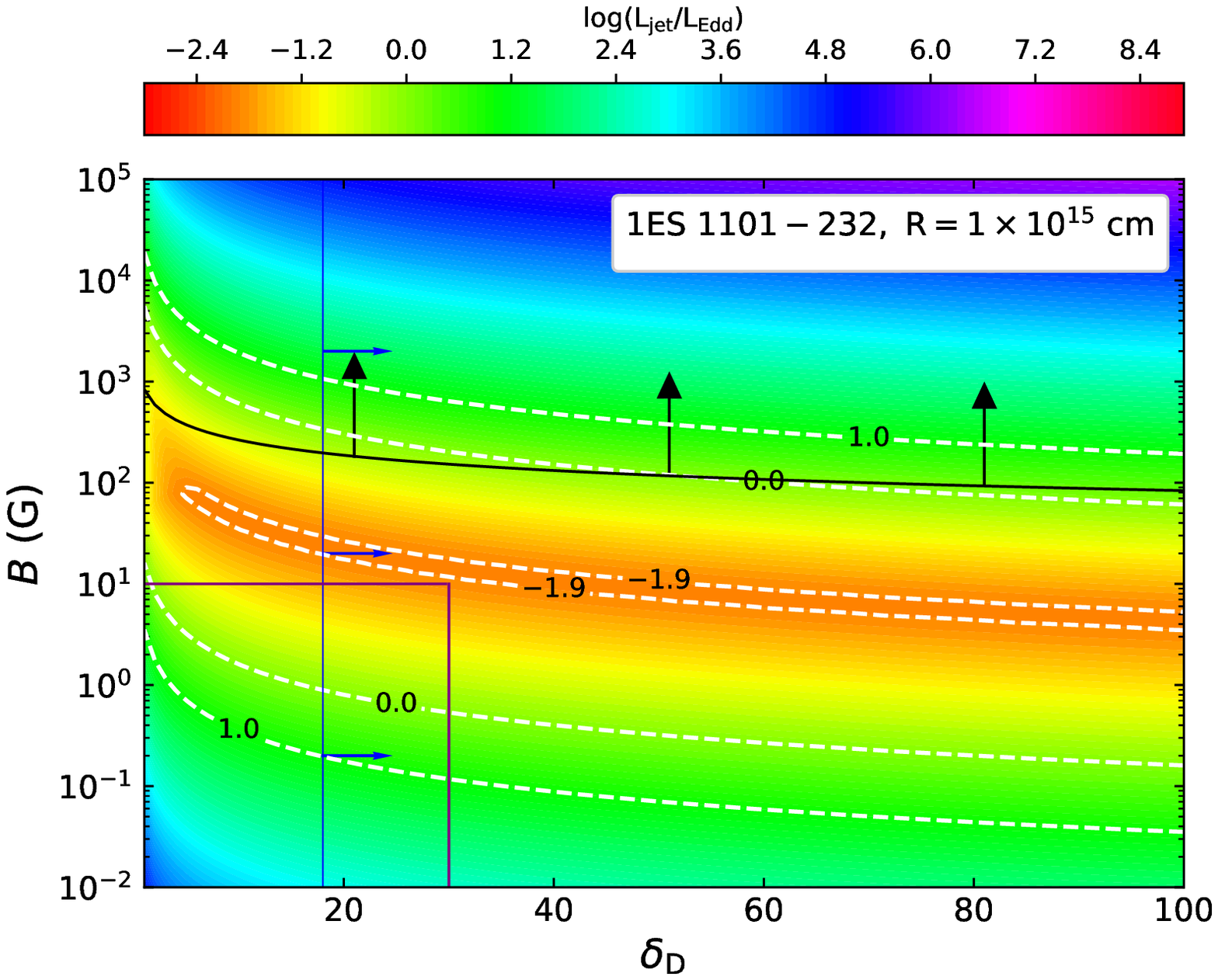}
}\hspace{-5mm}
\vspace{-1mm}
\quad
\subfloat{
\includegraphics[width=0.99\columnwidth]{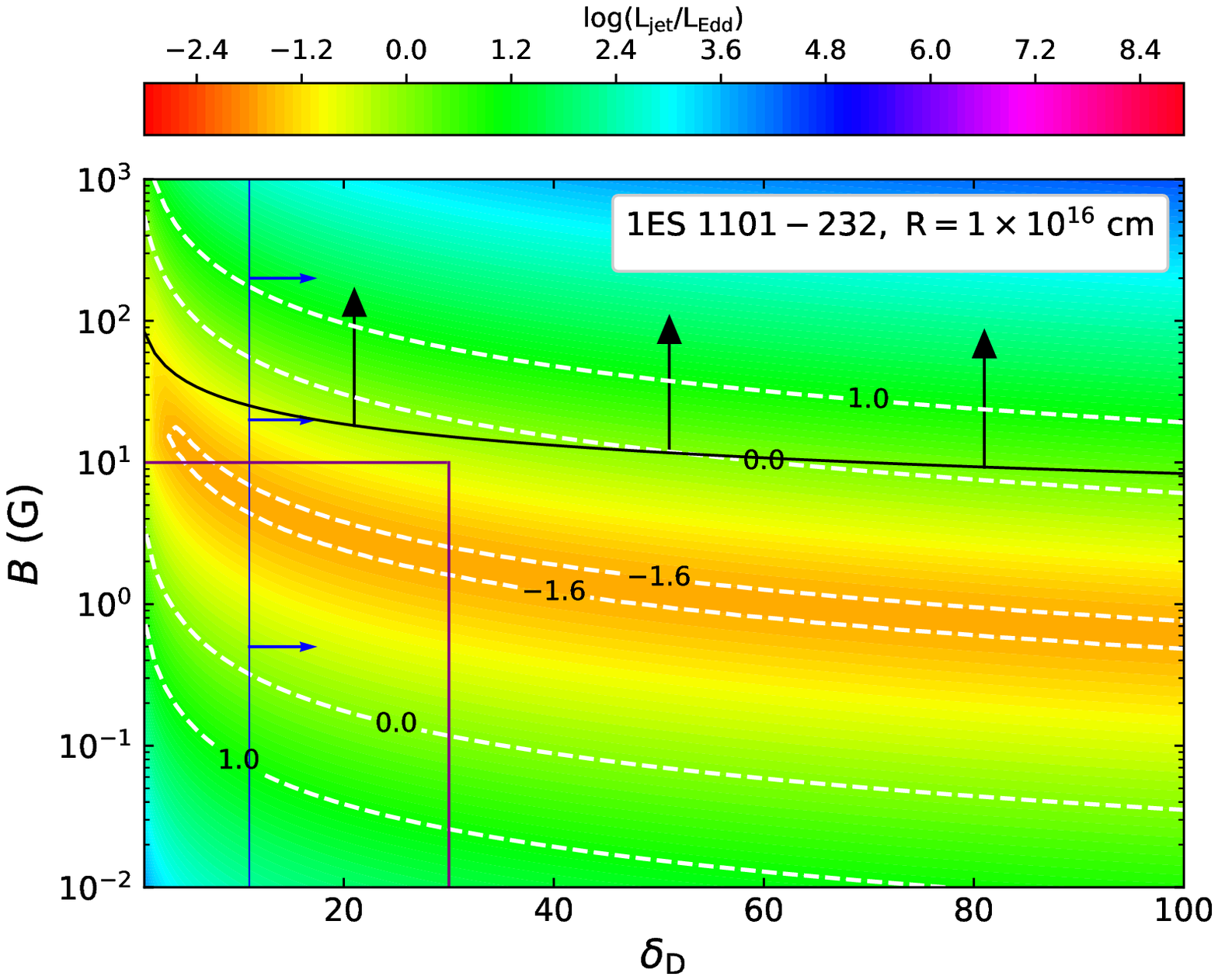}
}\hspace{-5mm}
\vspace{-1mm}
\quad
\subfloat{
\includegraphics[width=0.99\columnwidth]{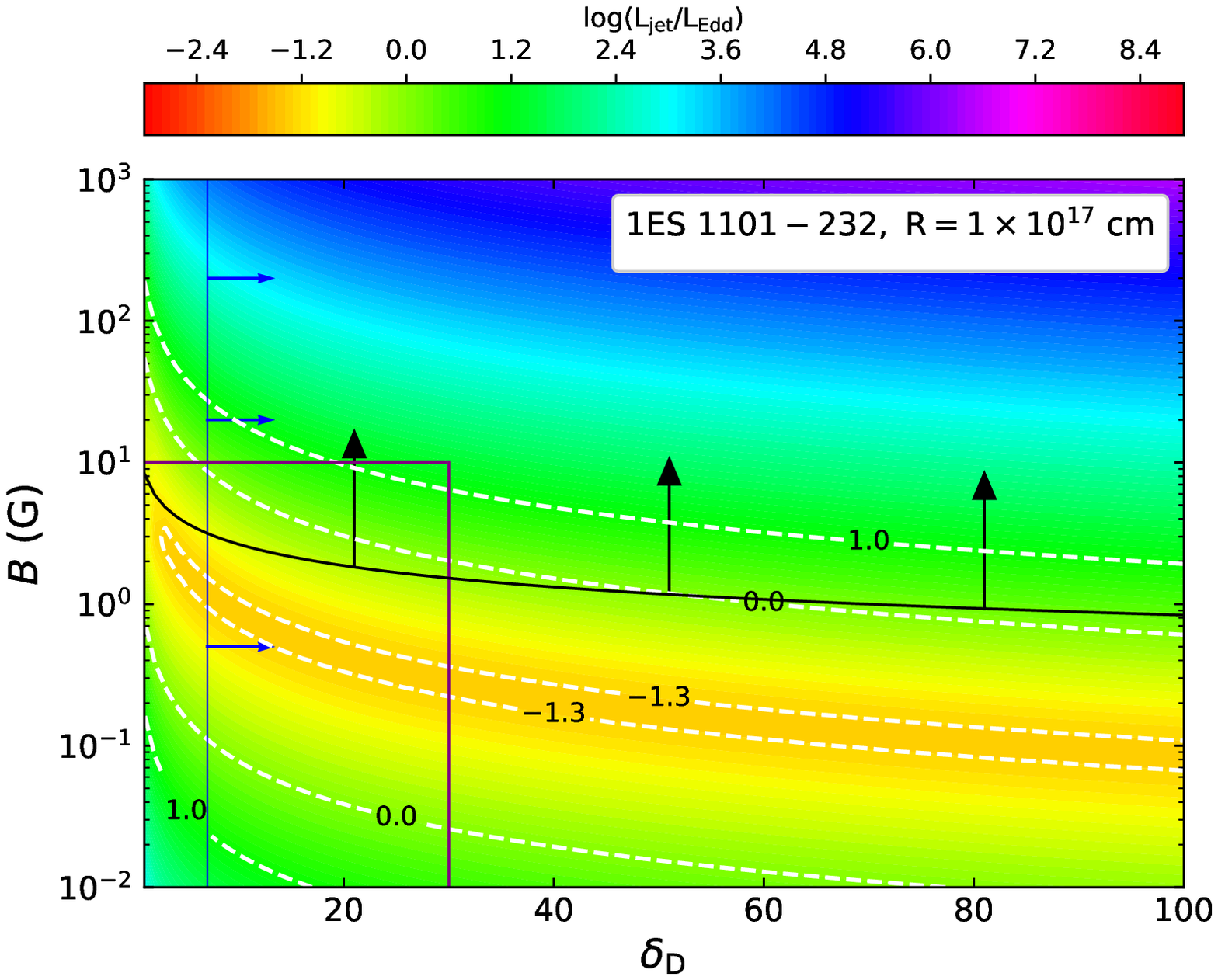}
}\hspace{-5mm}
\vspace{-1mm}
\quad
\subfloat{
\includegraphics[width=0.99\columnwidth]{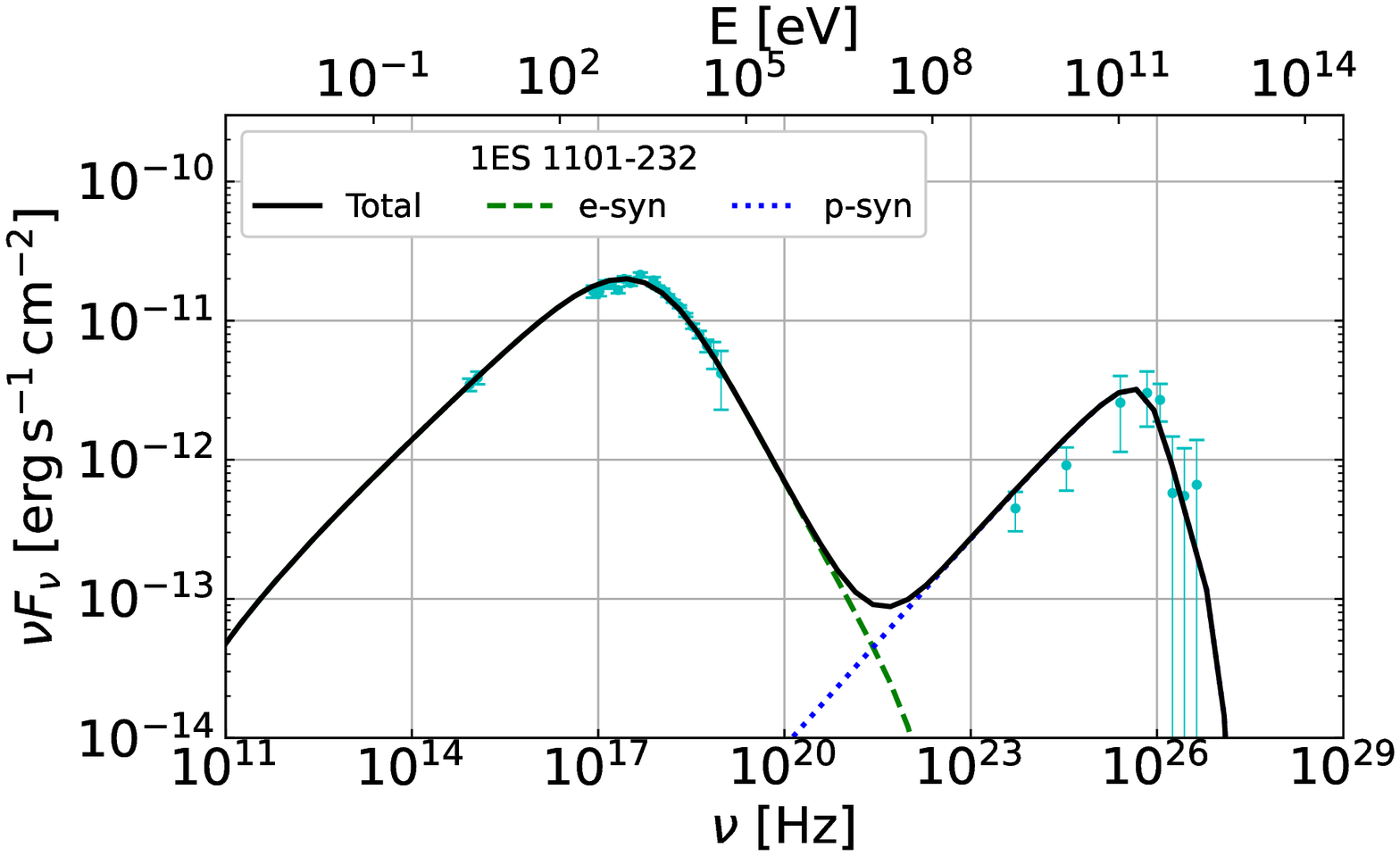}
}\hspace{-5mm}
\vspace{-1mm}
\quad
\subfloat{
\includegraphics[width=0.99\columnwidth]{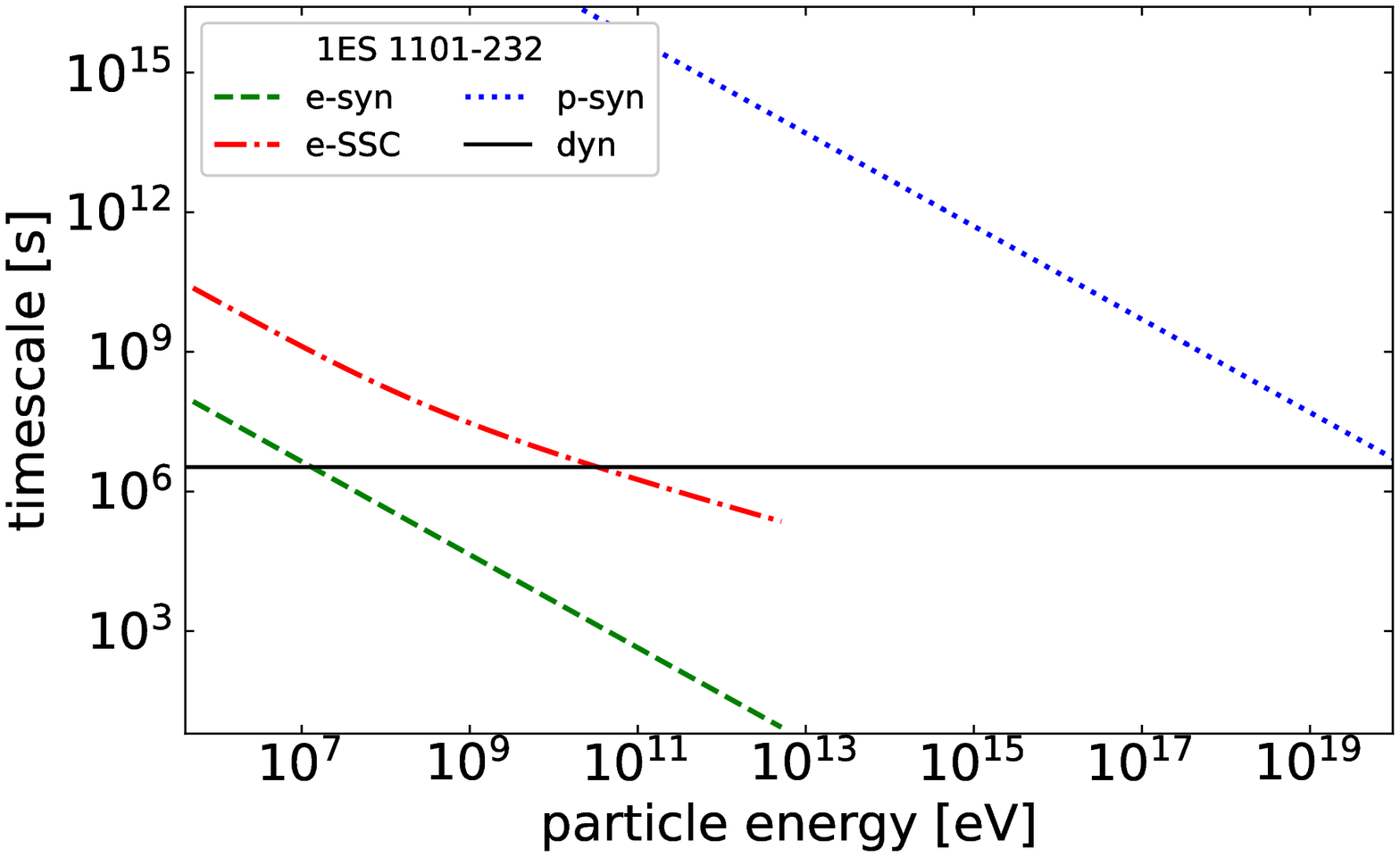}
}
\caption{Upper four panels: The ratio of $L_{\rm jet}/L_{\rm Edd}$ in the $\delta_{\rm D}$-$B$ diagram for 1ES 1101-232 peaking at 1.1~TeV, when setting $R=1\times 10^{14},~1\times 10^{15},~1\times 10^{16},~1\times 10^{17}~\rm cm$, respectively. Lower two panels: The fitting result of the SED of 1ES 1101-232 with the conventional one-zone model (left panel) and the corresponding timescales of various cooling processes for relativistic electrons and protons as a function of particle energy in the comoving frame (right panel). In the lower left panel, the cyan points from UV to GeV bands and in the VHE band are quasi-simultaneous data taken from \cite{2018MNRAS.477.4257C} and archival data taken from \cite{2007AA...470..475A}, respectively. The parameters are the same as those shown in Table~\ref{parameters}. The line styles in all panels have the same meaning as in Fig.~\ref{0229}. \label{1101}}
\end{figure*}

\begin{figure*}
\centering
\subfloat{
\includegraphics[width=1\columnwidth]{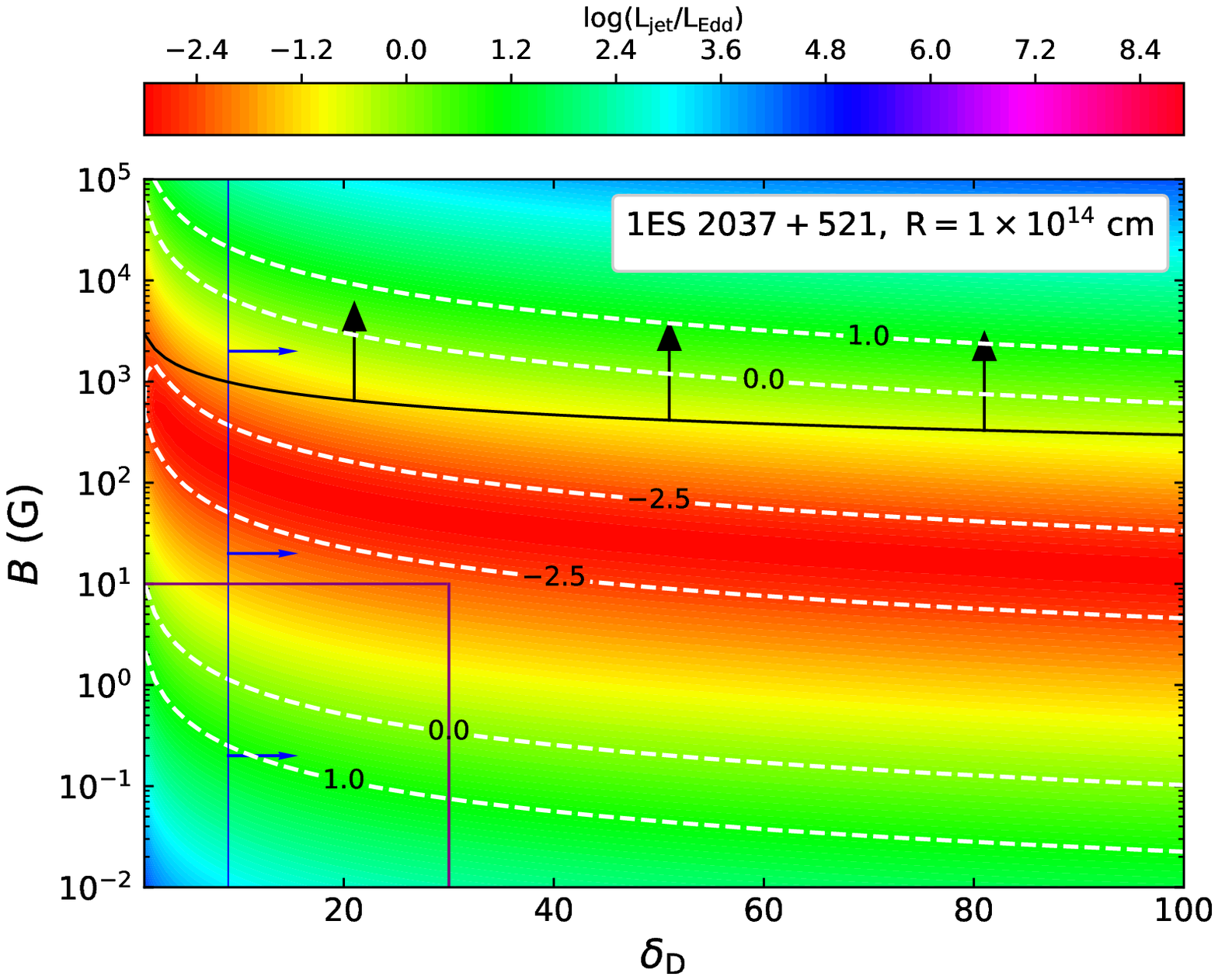}
}\hspace{-5mm}
\vspace{-1mm}
\quad
\subfloat{
\includegraphics[width=1\columnwidth]{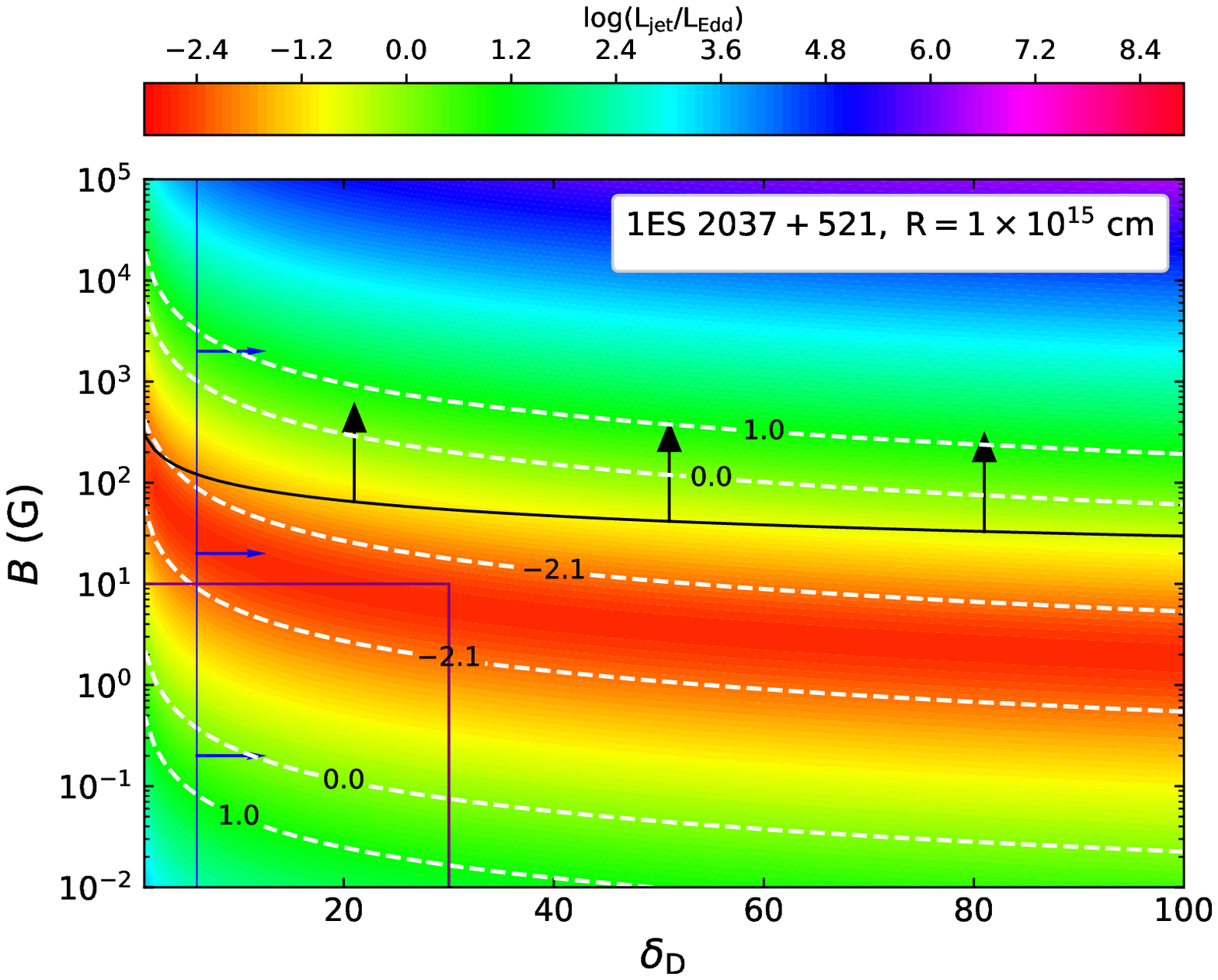}
}\hspace{-5mm}
\vspace{-1mm}
\quad
\subfloat{
\includegraphics[width=0.99\columnwidth]{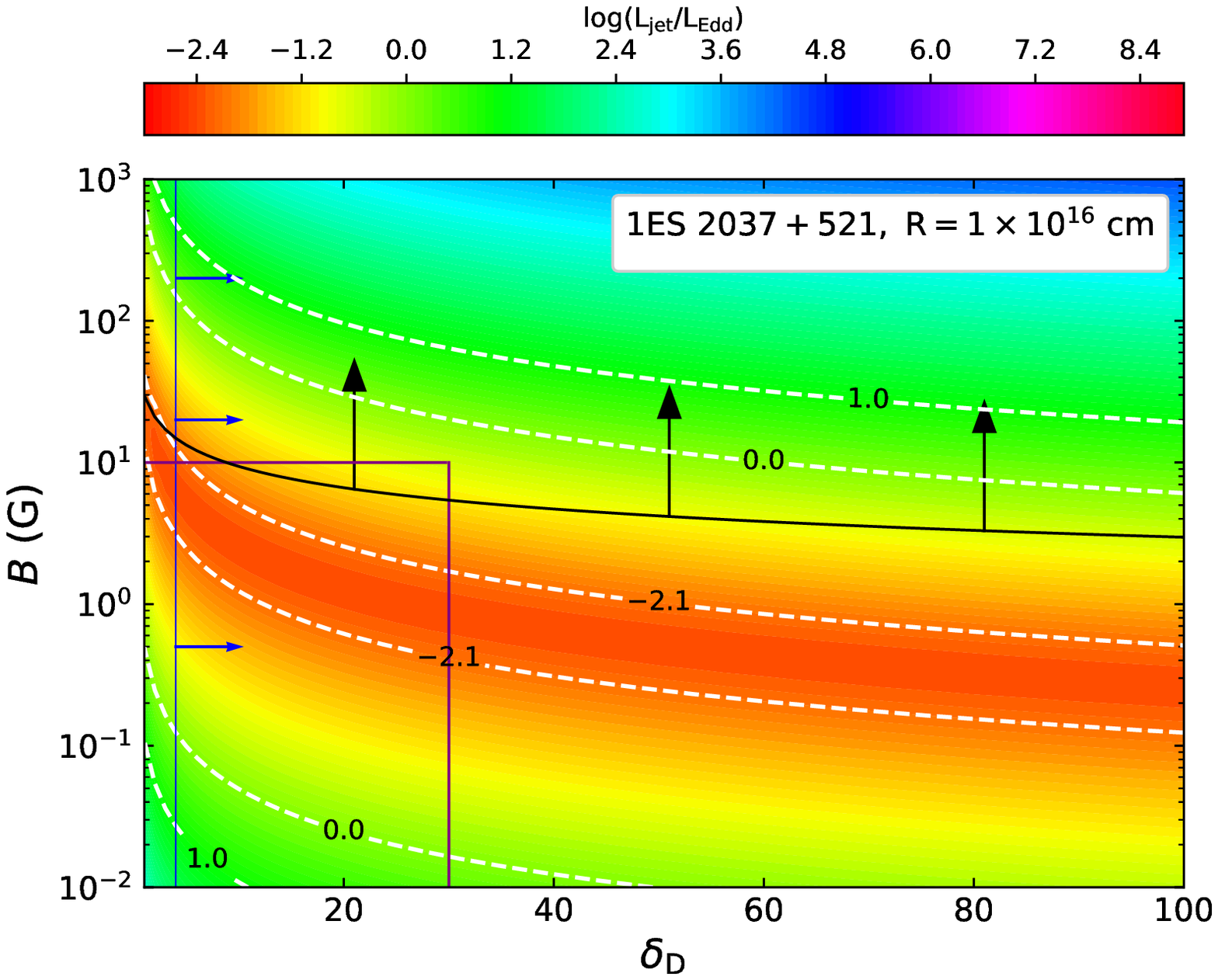}
}\hspace{-5mm}
\vspace{-1mm}
\quad
\subfloat{
\includegraphics[width=0.99\columnwidth]{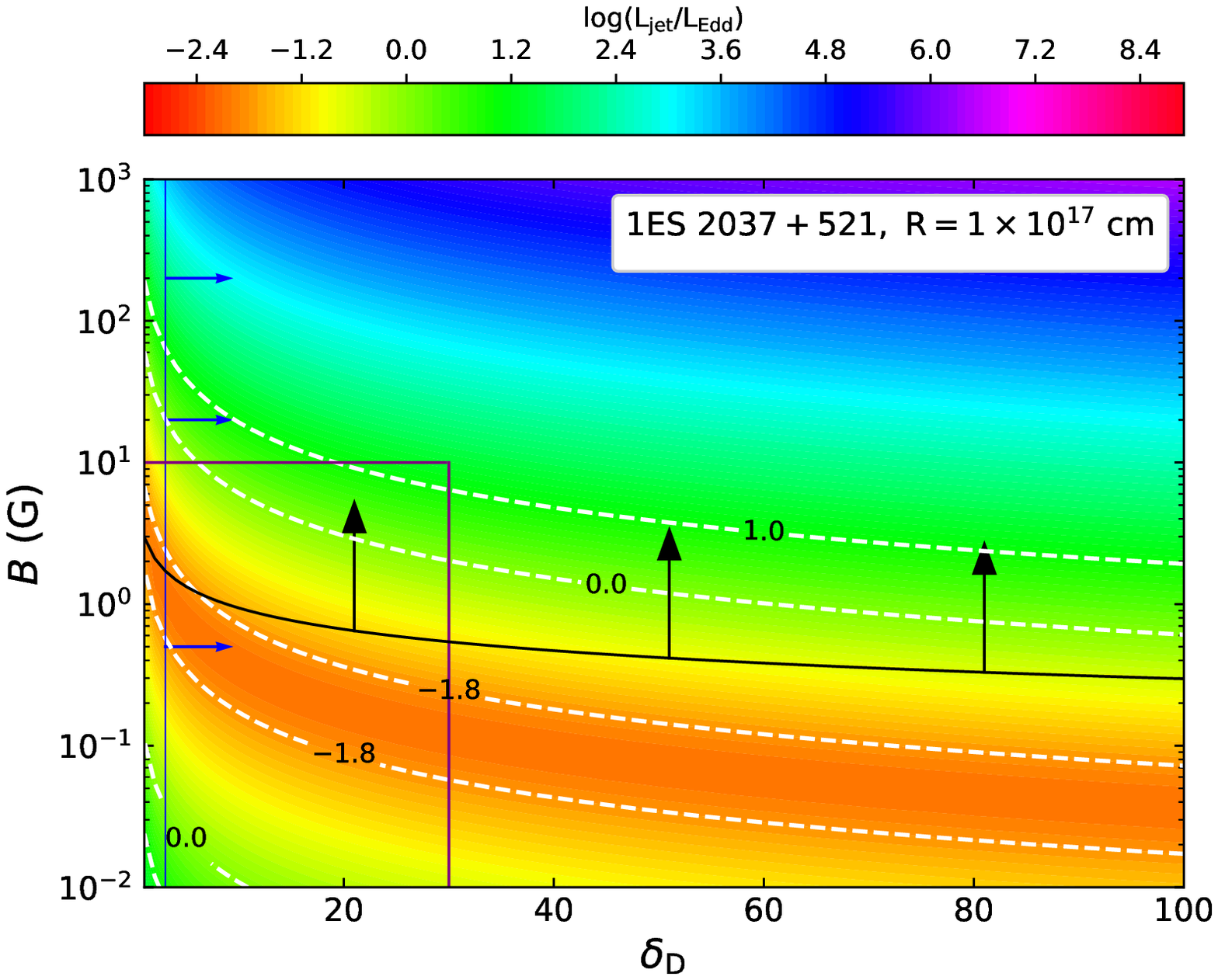}
}\hspace{-5mm}
\vspace{-1mm}
\quad
\subfloat{
\includegraphics[width=0.99\columnwidth]{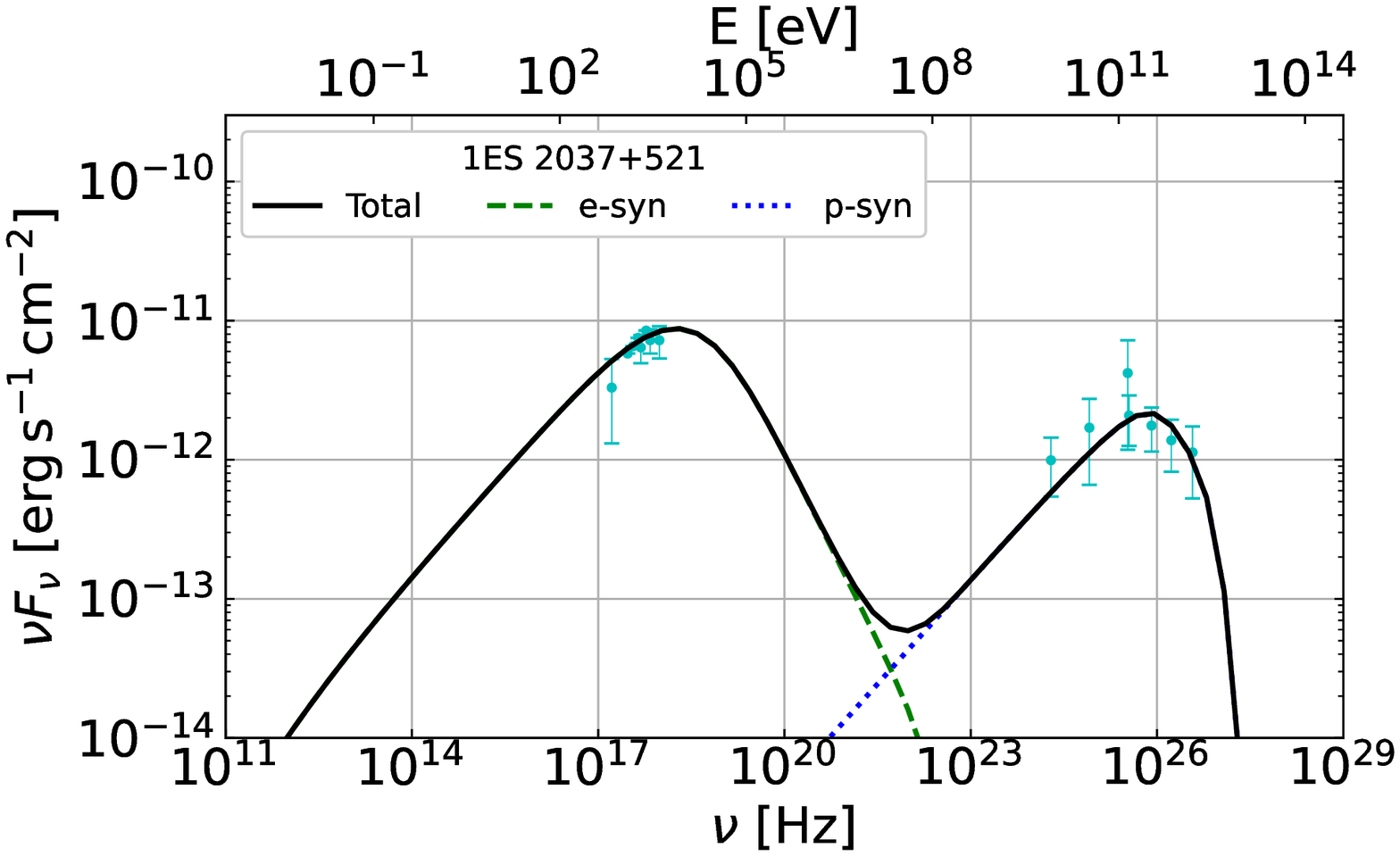}
}\hspace{-5mm}
\vspace{-1mm}
\quad
\subfloat{
\includegraphics[width=0.99\columnwidth]{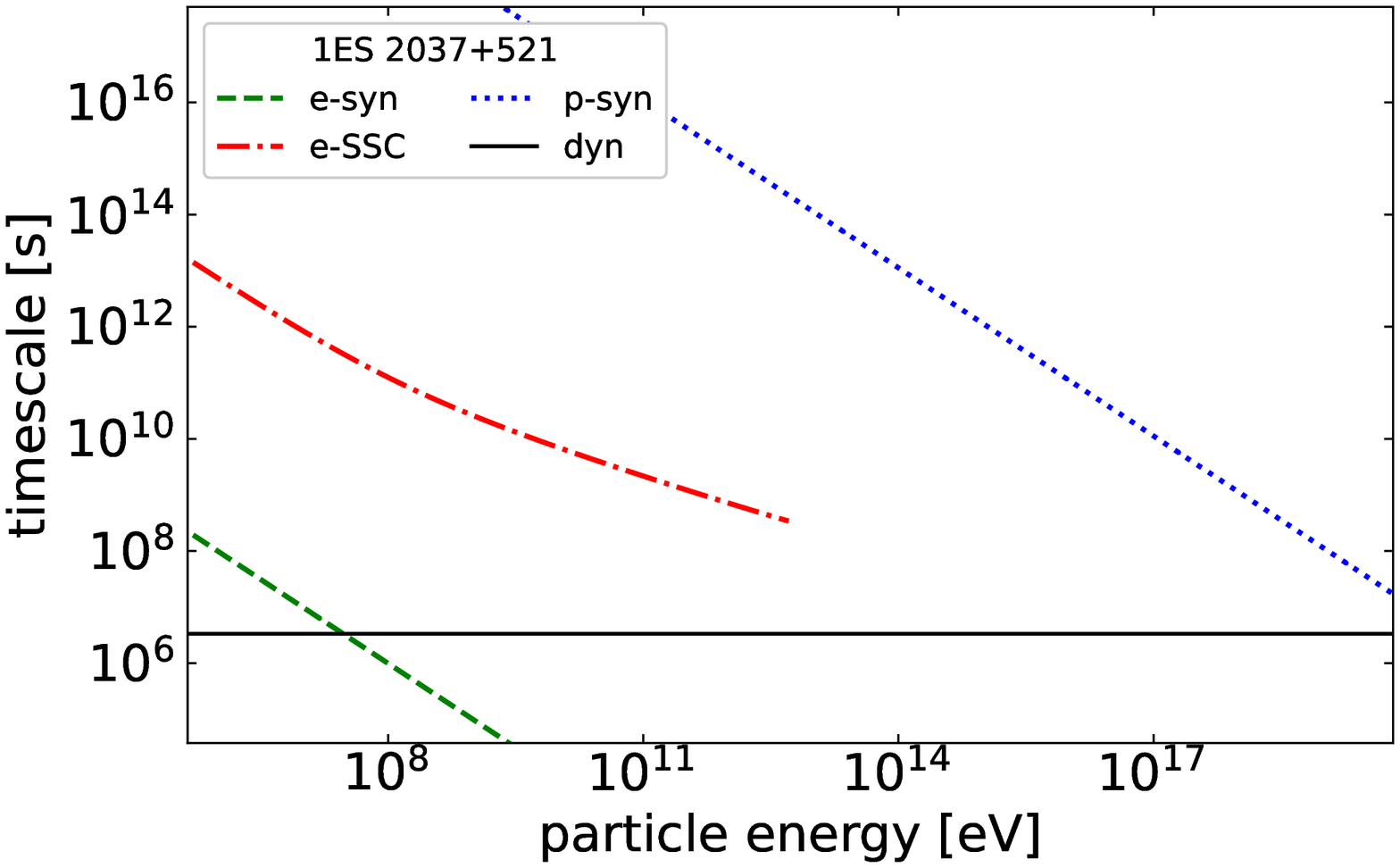}
}
\caption{Upper four panels: The ratio of $L_{\rm jet}/L_{\rm Edd}$ in the $\delta_{\rm D}$-$B$ diagram for 1ES 2037+521 peaking at 140~GeV, when setting $R=1\times 10^{14},~1\times 10^{15},~1\times 10^{16},~1\times 10^{17}~\rm cm$, respectively. Lower two panels: The fitting result of the quasi-simultaneous SED of 1ES 2037+521 with the conventional one-zone model (left panel) and the corresponding timescales of various cooling processes for relativistic electrons and protons as a function of particle energy in the comoving frame (right panel). In the lower left panel, the cyan points are data taken from \cite{2020ApJS..247...16A}. The parameters are the same as those shown in Table~\ref{parameters}. The line styles in all panels have the same meaning as in Fig.~\ref{0229}. \label{2037}}
\end{figure*}

\begin{figure*}
\centering
\subfloat{
\includegraphics[width=1\columnwidth]{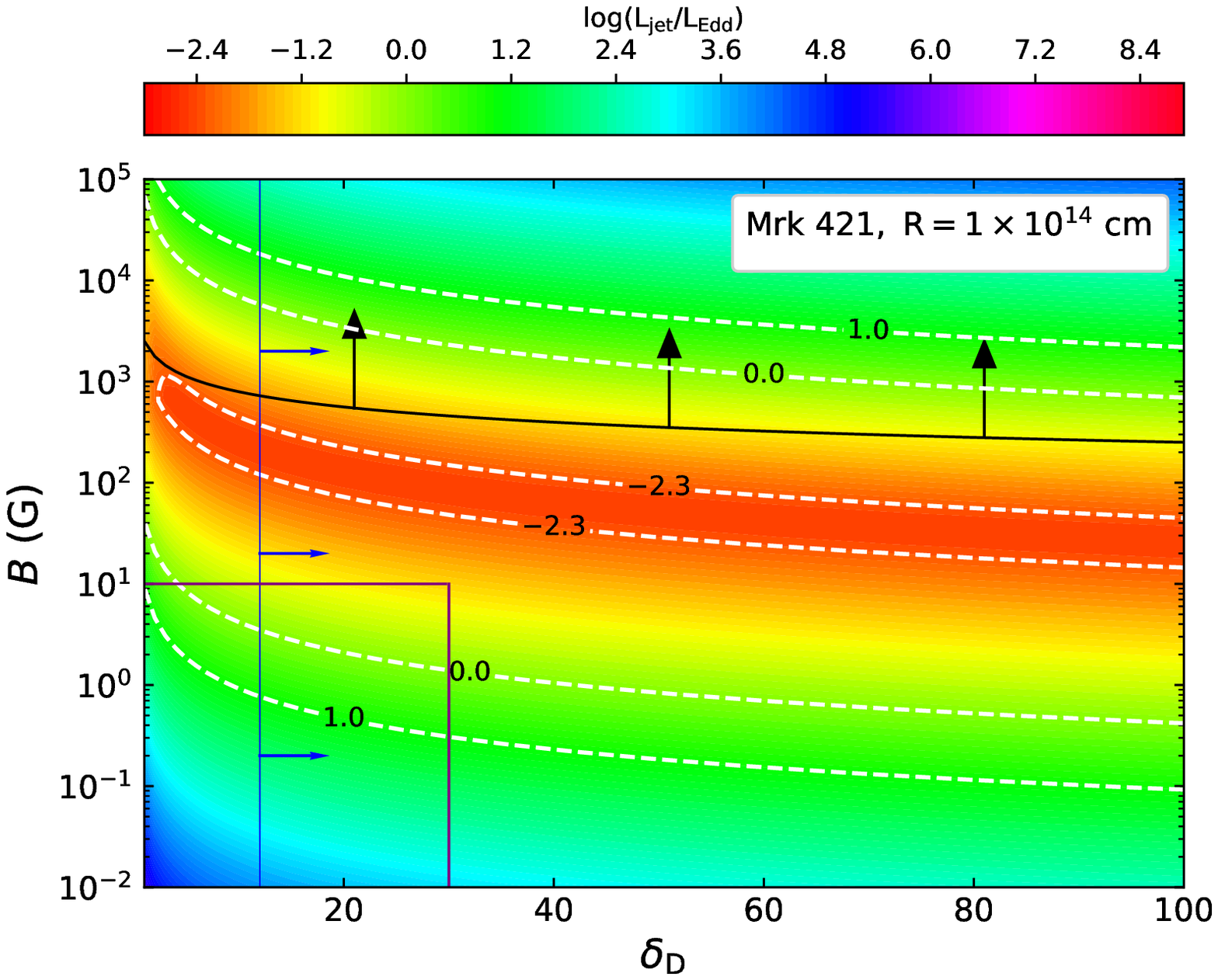}
}\hspace{-5mm}
\vspace{-1mm}
\quad
\subfloat{
\includegraphics[width=1\columnwidth]{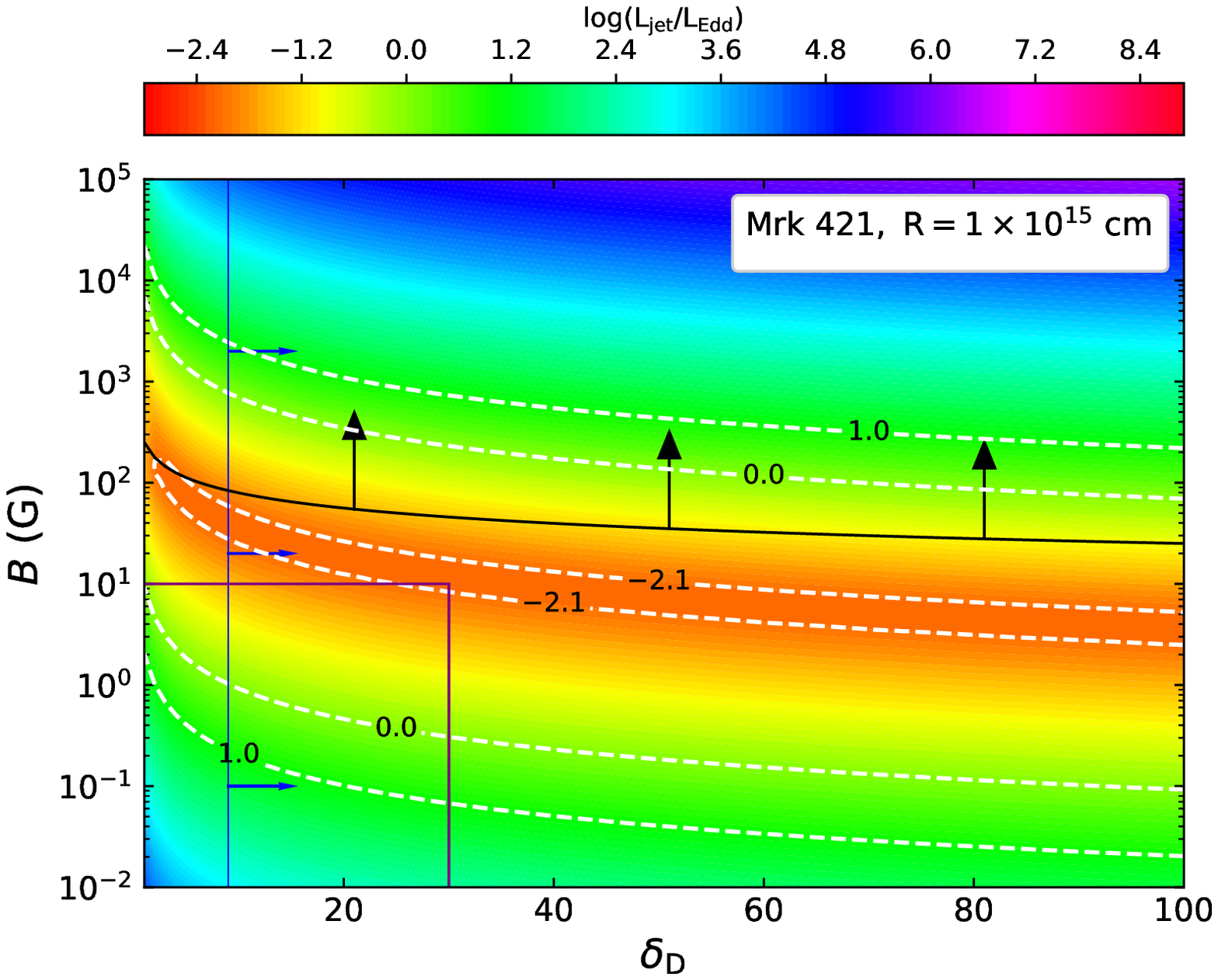}
}\hspace{-5mm}
\vspace{-1mm}
\quad
\subfloat{
\includegraphics[width=0.99\columnwidth]{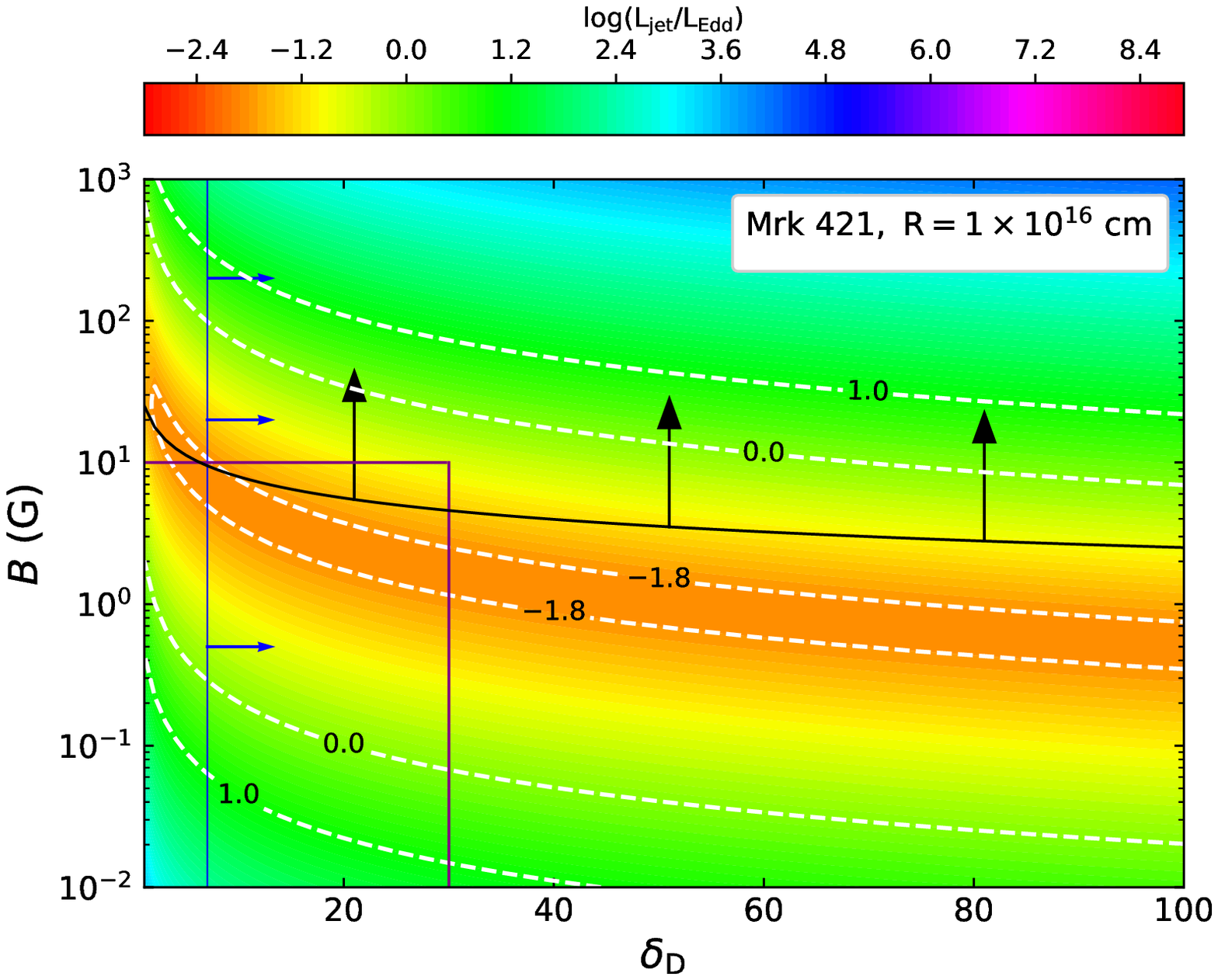}
}\hspace{-5mm}
\vspace{-1mm}
\quad
\subfloat{
\includegraphics[width=0.99\columnwidth]{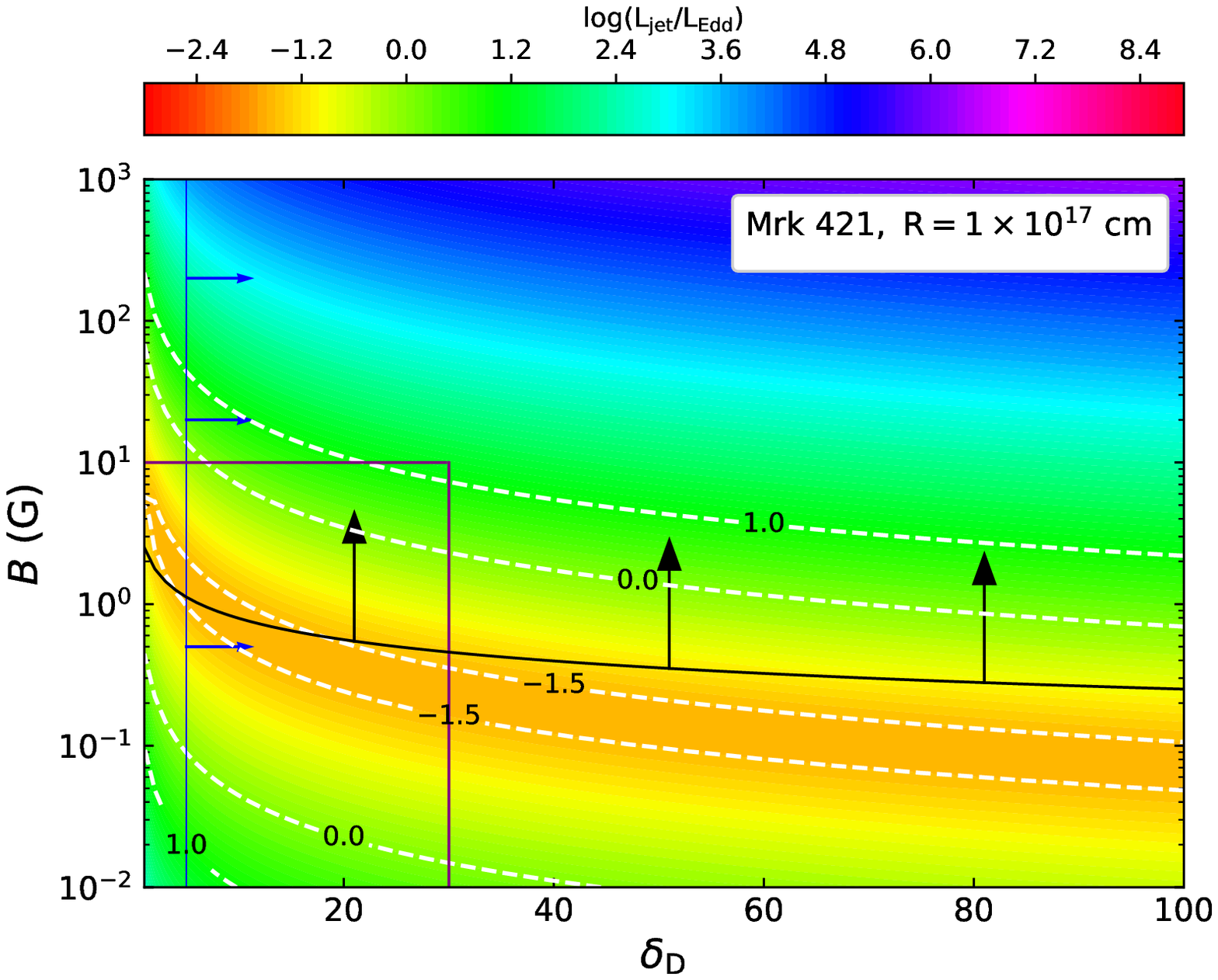}
}\hspace{-5mm}
\vspace{-1mm}
\quad
\subfloat{
\includegraphics[width=0.99\columnwidth]{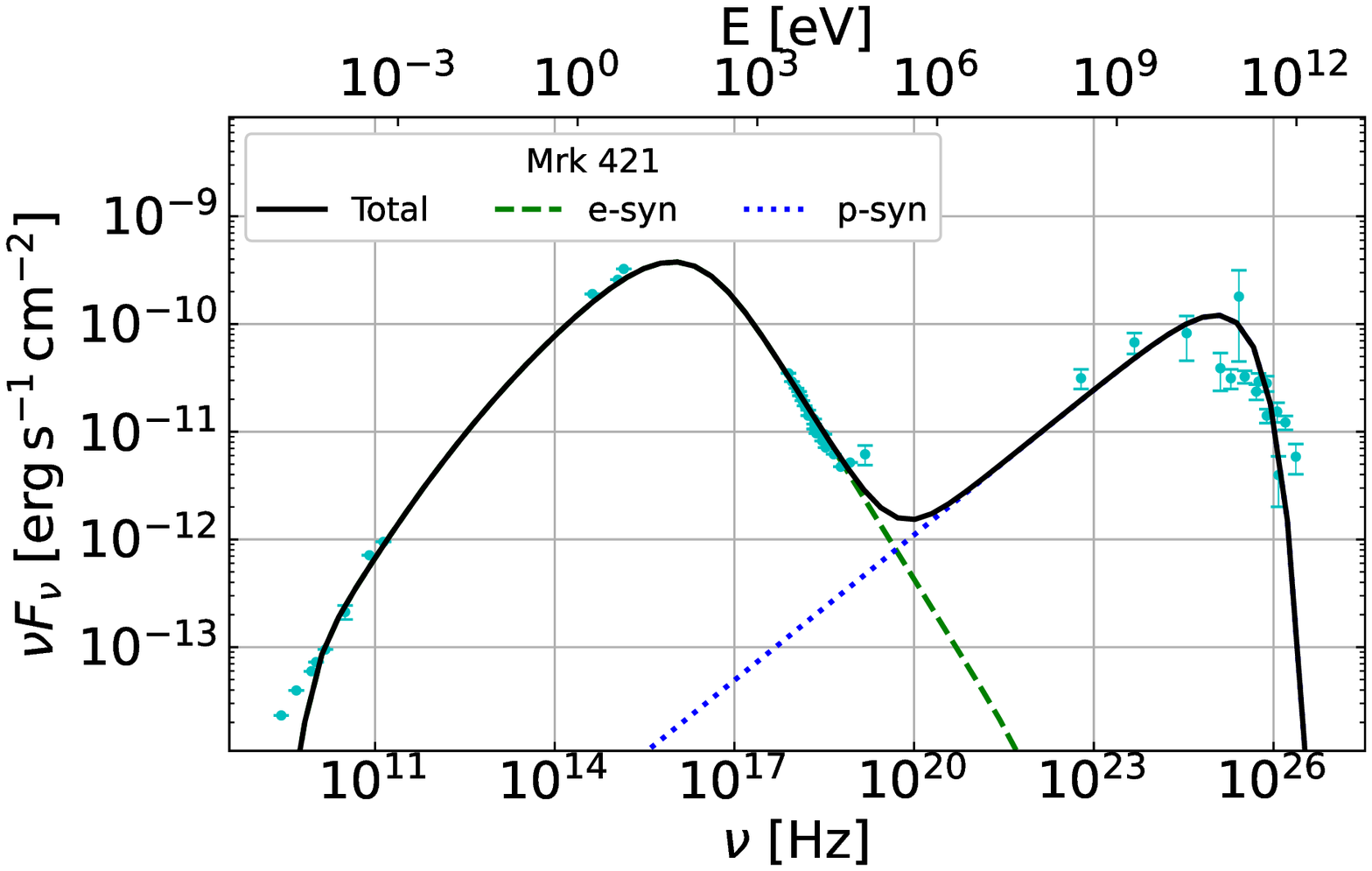}
}\hspace{-5mm}
\vspace{-1mm}
\quad
\subfloat{
\includegraphics[width=0.99\columnwidth]{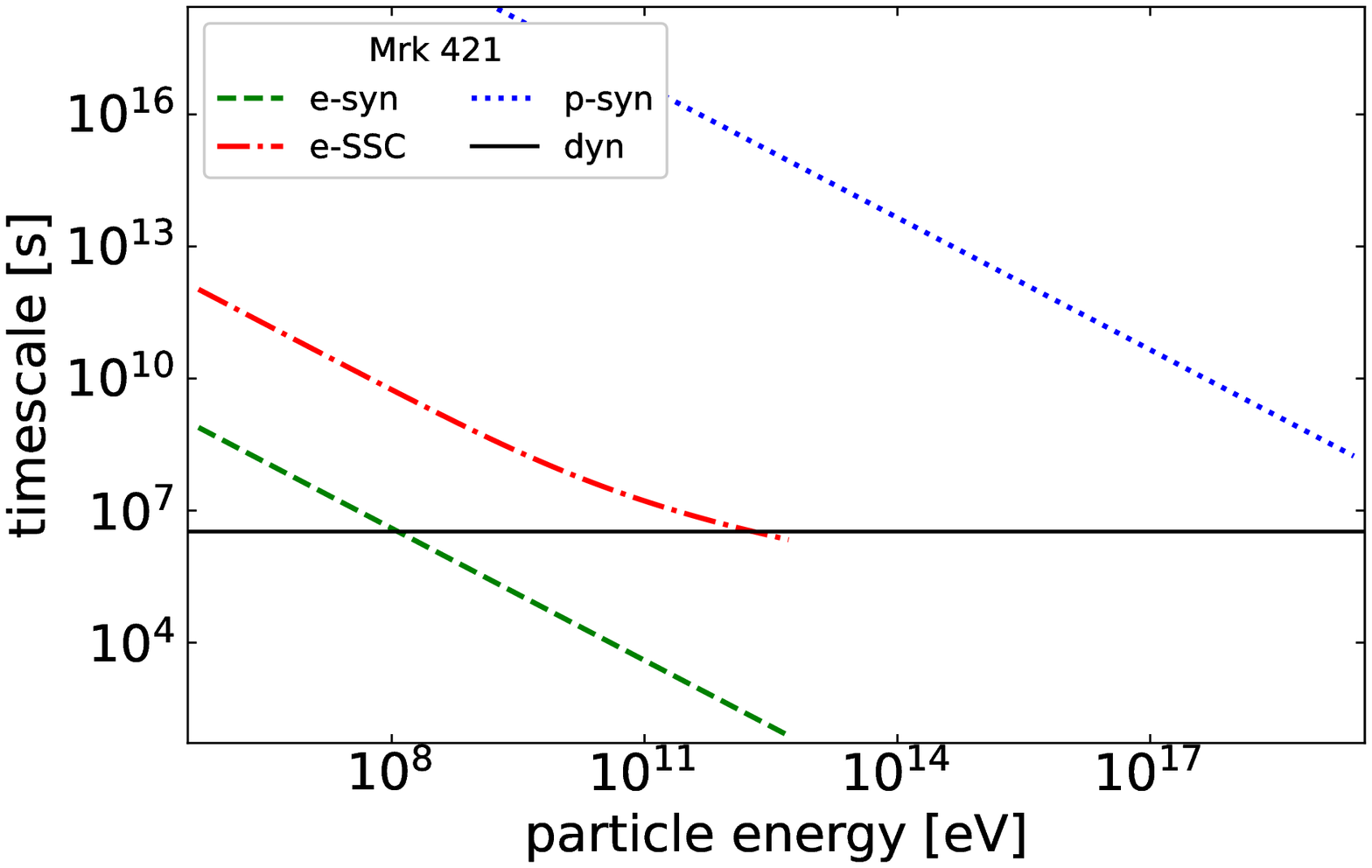}
}
\caption{Upper four panels: The ratio of $L_{\rm jet}/L_{\rm Edd}$ in the $\delta_{\rm D}$-$B$ diagram for Mrk 421 peaking at 50~GeV, when setting $R=1\times 10^{14},~1\times 10^{15},~1\times 10^{16},~1\times 10^{17}~\rm cm$, respectively. Lower two panels: The fitting result of the quasi-simultaneous SED of Mrk 421 with the conventional one-zone model (left panel) and the corresponding timescales of various cooling processes for relativistic electrons and protons as a function of particle energy in the comoving frame (right panel). In the lower left panel, the cyan points are data taken from \cite{2016ApJ...827...55K}. The parameters are the same as those shown in Table~\ref{parameters}. The line styles in all panels have the same meaning as in Fig.~\ref{0229}. \label{421-0.1T}}
\end{figure*}

\begin{figure*}
\centering
\subfloat{
\includegraphics[width=1\columnwidth]{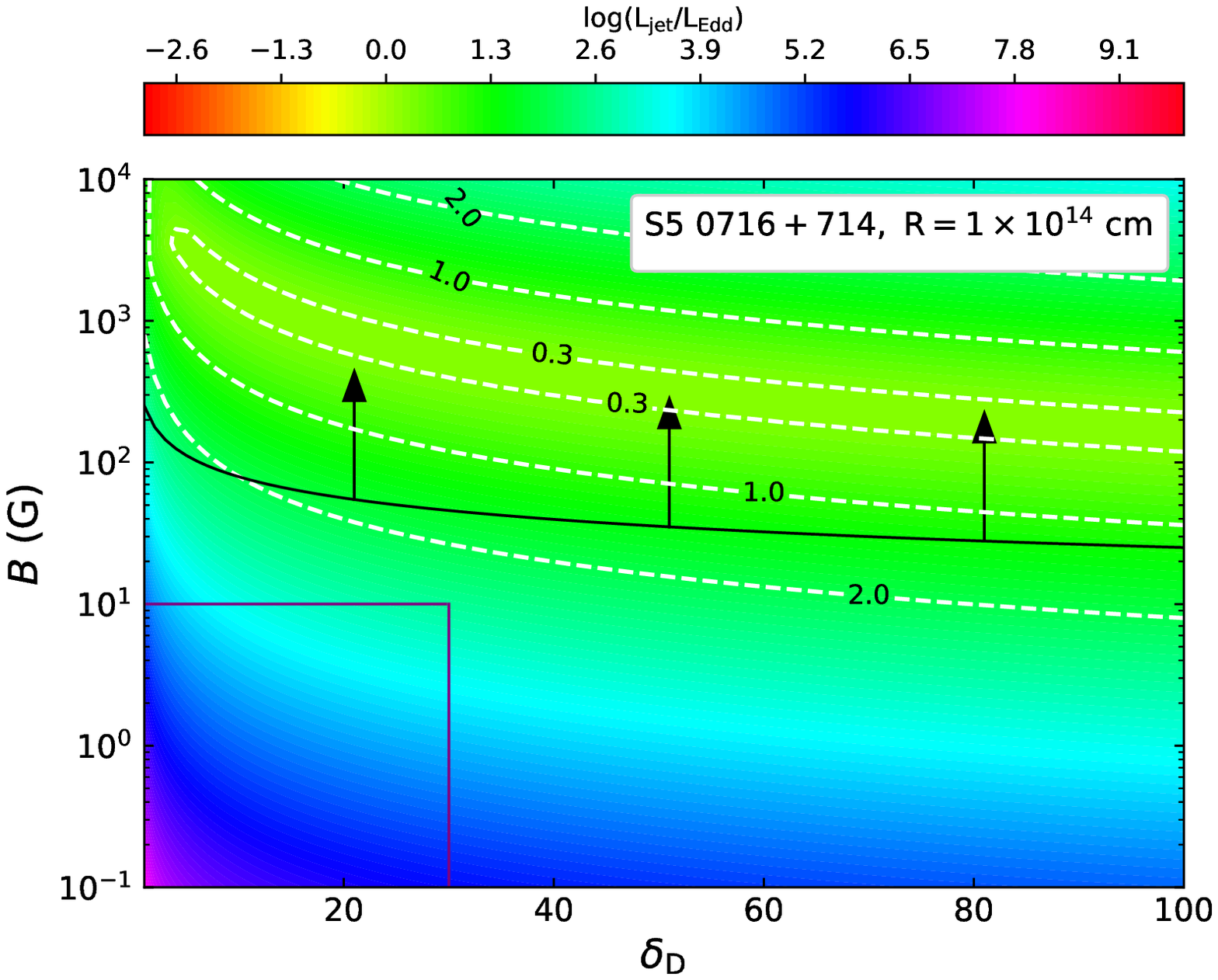}
}\hspace{-5mm}
\vspace{-1mm}
\quad
\subfloat{
\includegraphics[width=1\columnwidth]{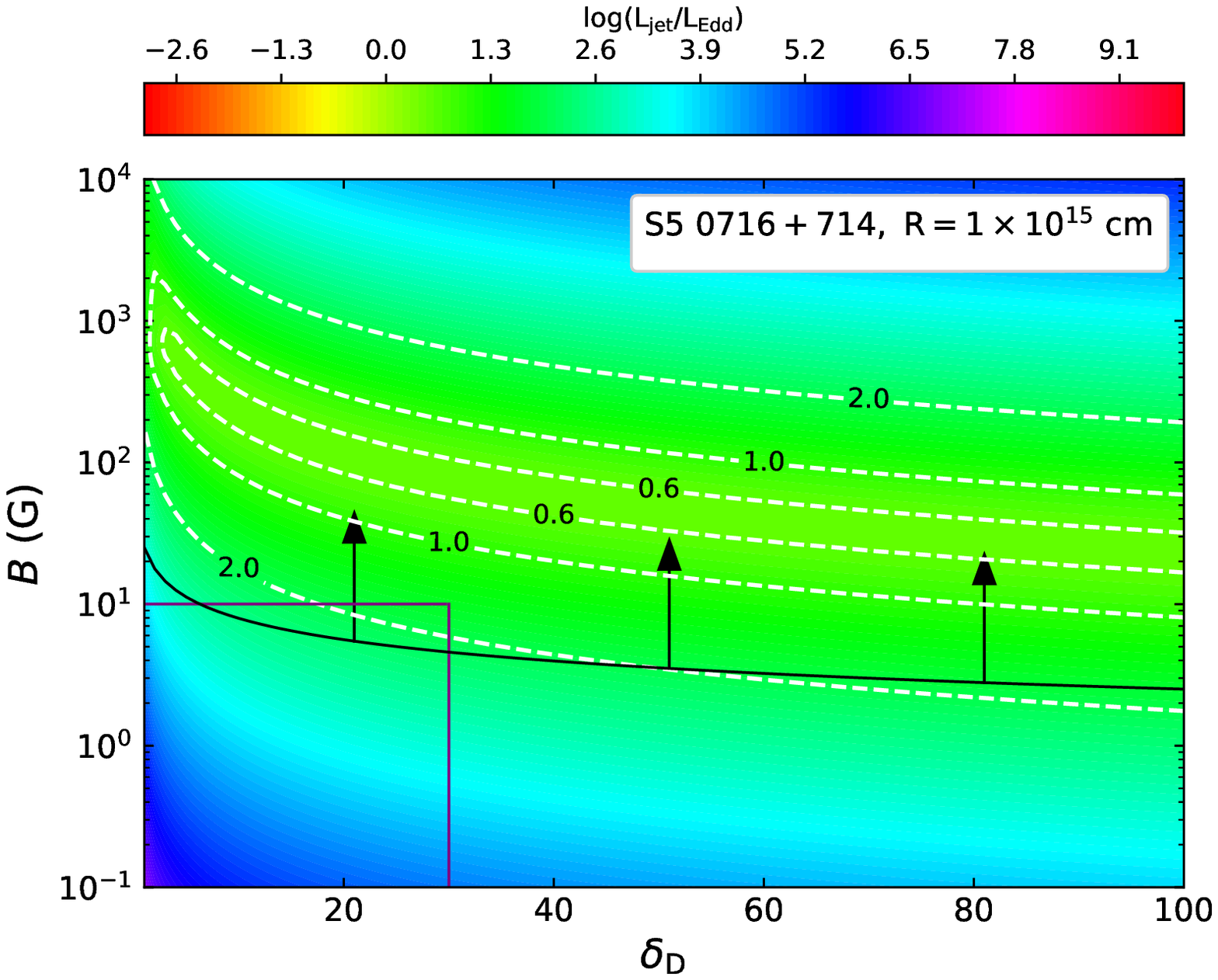}
}\hspace{-5mm}
\vspace{-1mm}
\quad
\subfloat{
\includegraphics[width=0.99\columnwidth]{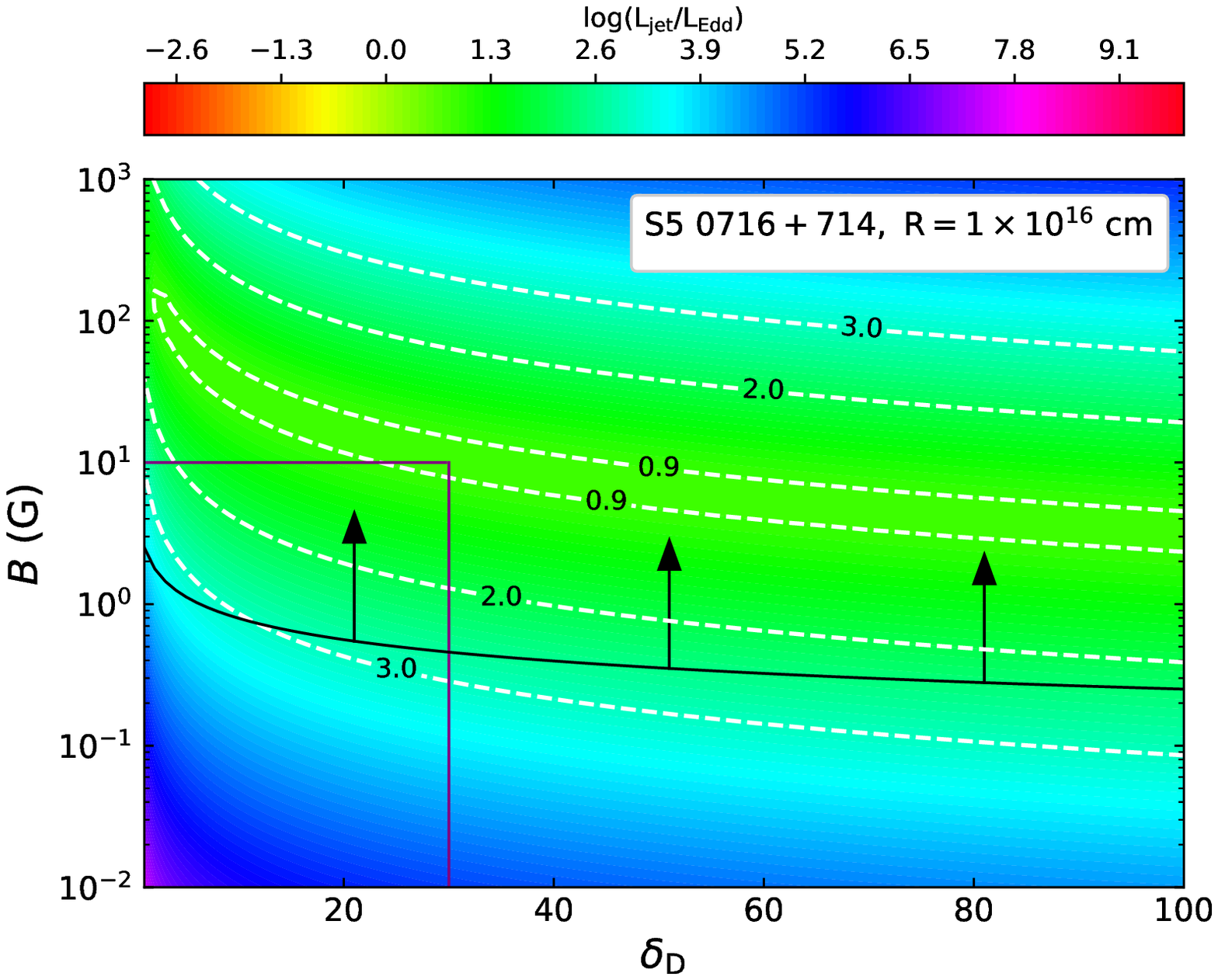}
}\hspace{-5mm}
\vspace{-1mm}
\quad
\subfloat{
\includegraphics[width=0.99\columnwidth]{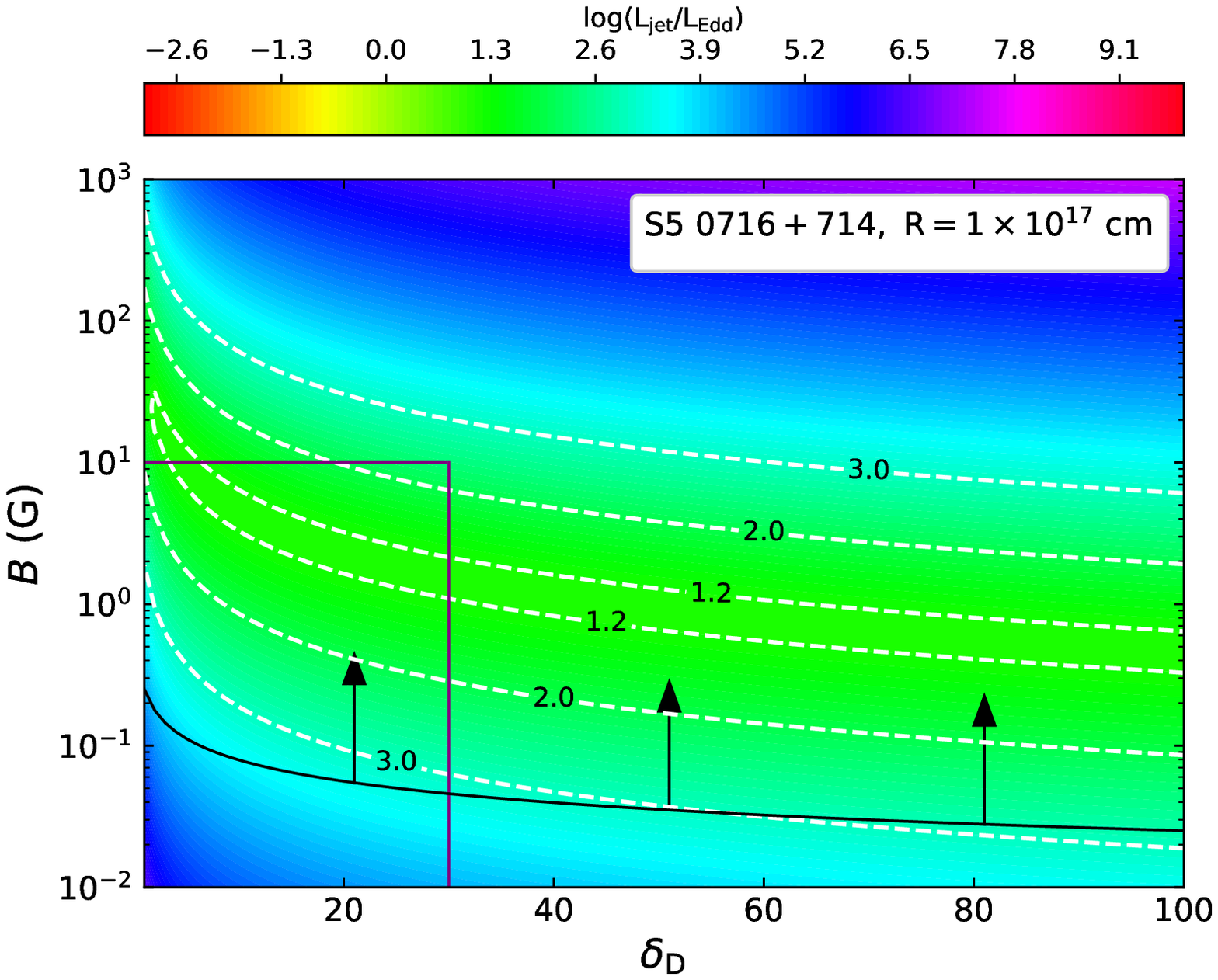}
}
\caption{The ratio of $L_{\rm jet}/L_{\rm Edd}$ in the $\delta_{\rm D}$-$B$ diagram for S5 0716+714 peaking at 1~GeV, when setting $R=1\times 10^{14},~1\times 10^{15},~1\times 10^{16},~1\times 10^{17}~\rm cm$, respectively. The line styles in all panels have the same meaning as in Fig.~\ref{0229}. \label{S5}}
\end{figure*}

\begin{figure*}
\centering
\subfloat{
\includegraphics[width=1\columnwidth]{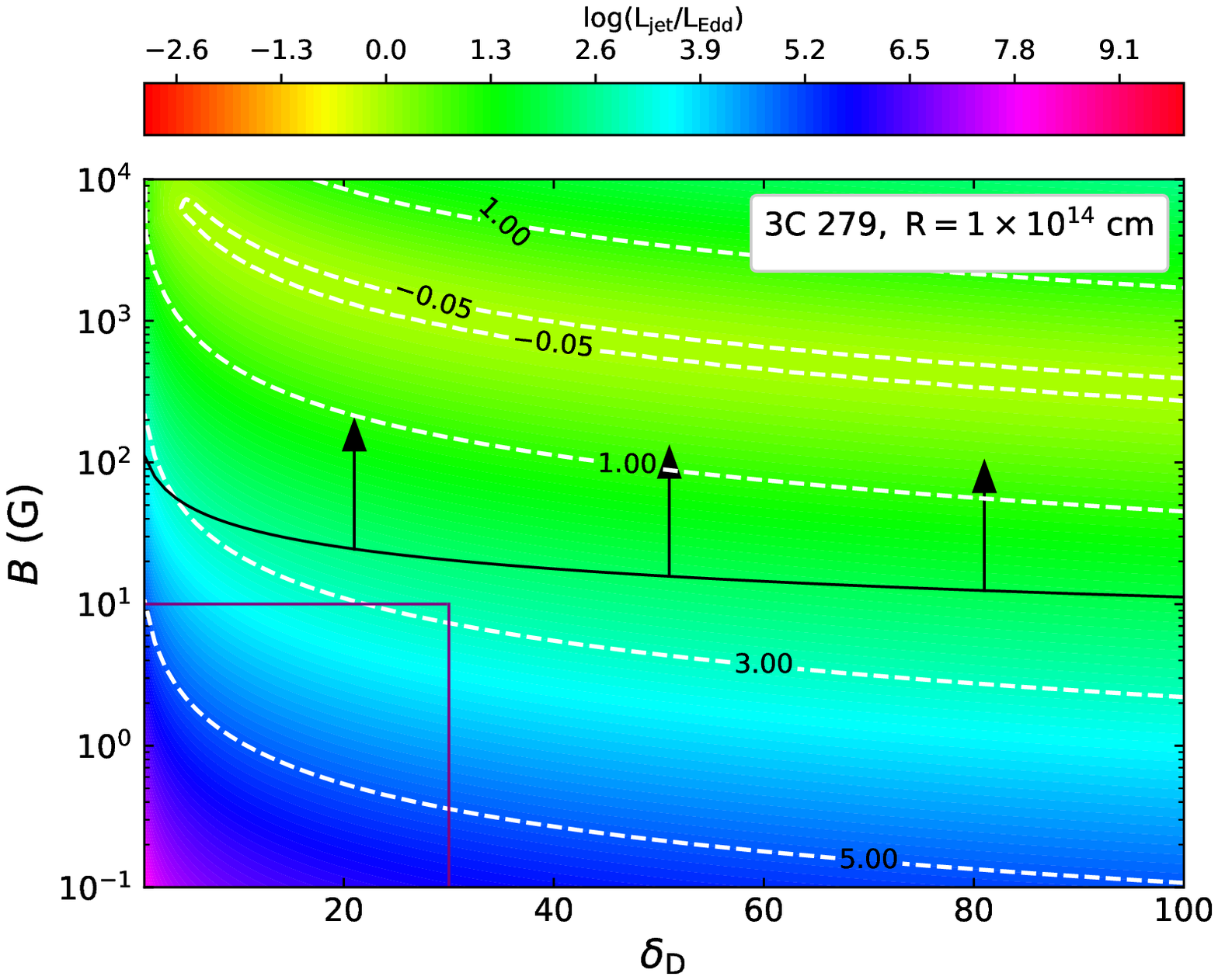}
}\hspace{-5mm}
\vspace{-1mm}
\quad
\subfloat{
\includegraphics[width=1\columnwidth]{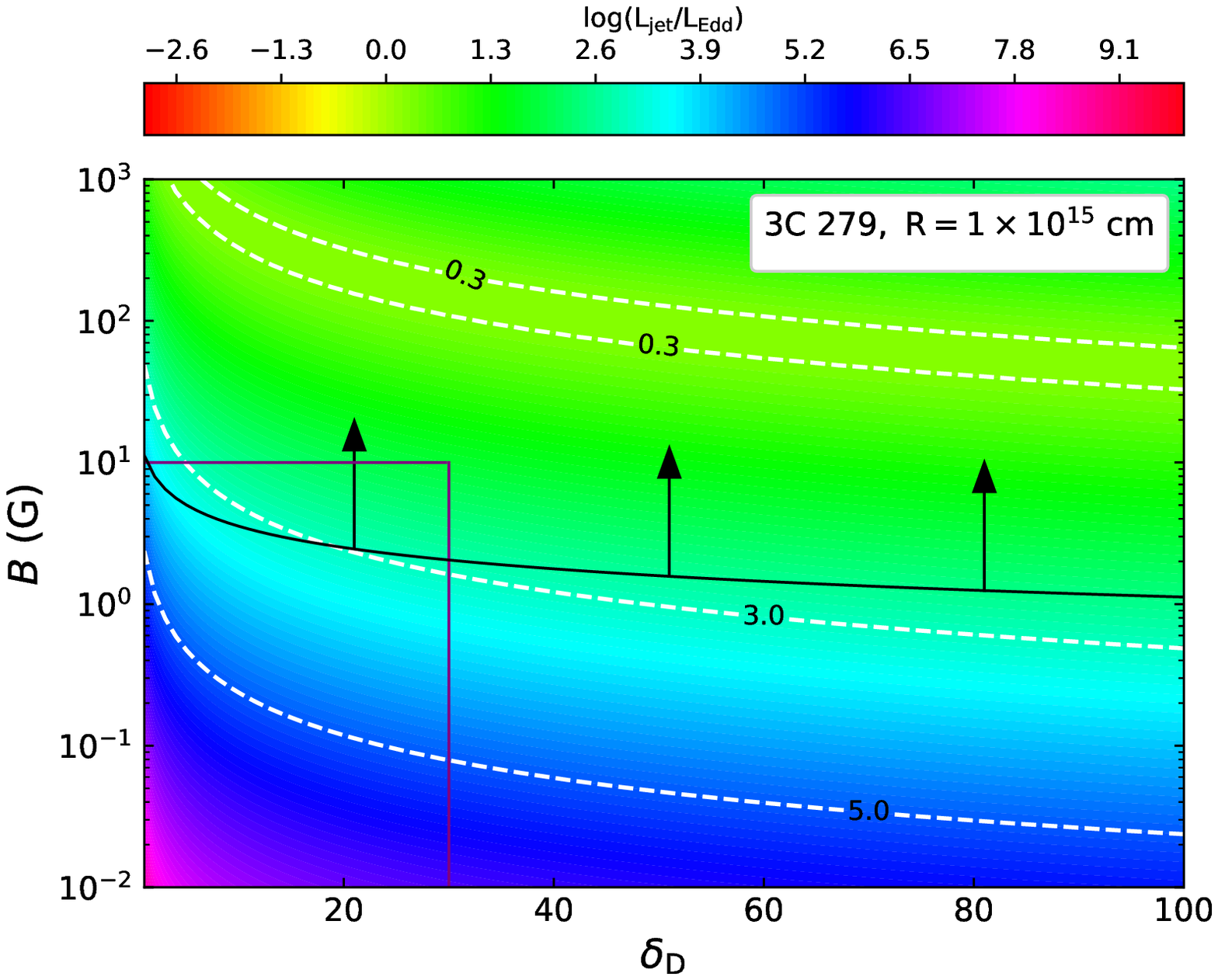}
}\hspace{-5mm}
\vspace{-1mm}
\quad
\subfloat{
\includegraphics[width=0.99\columnwidth]{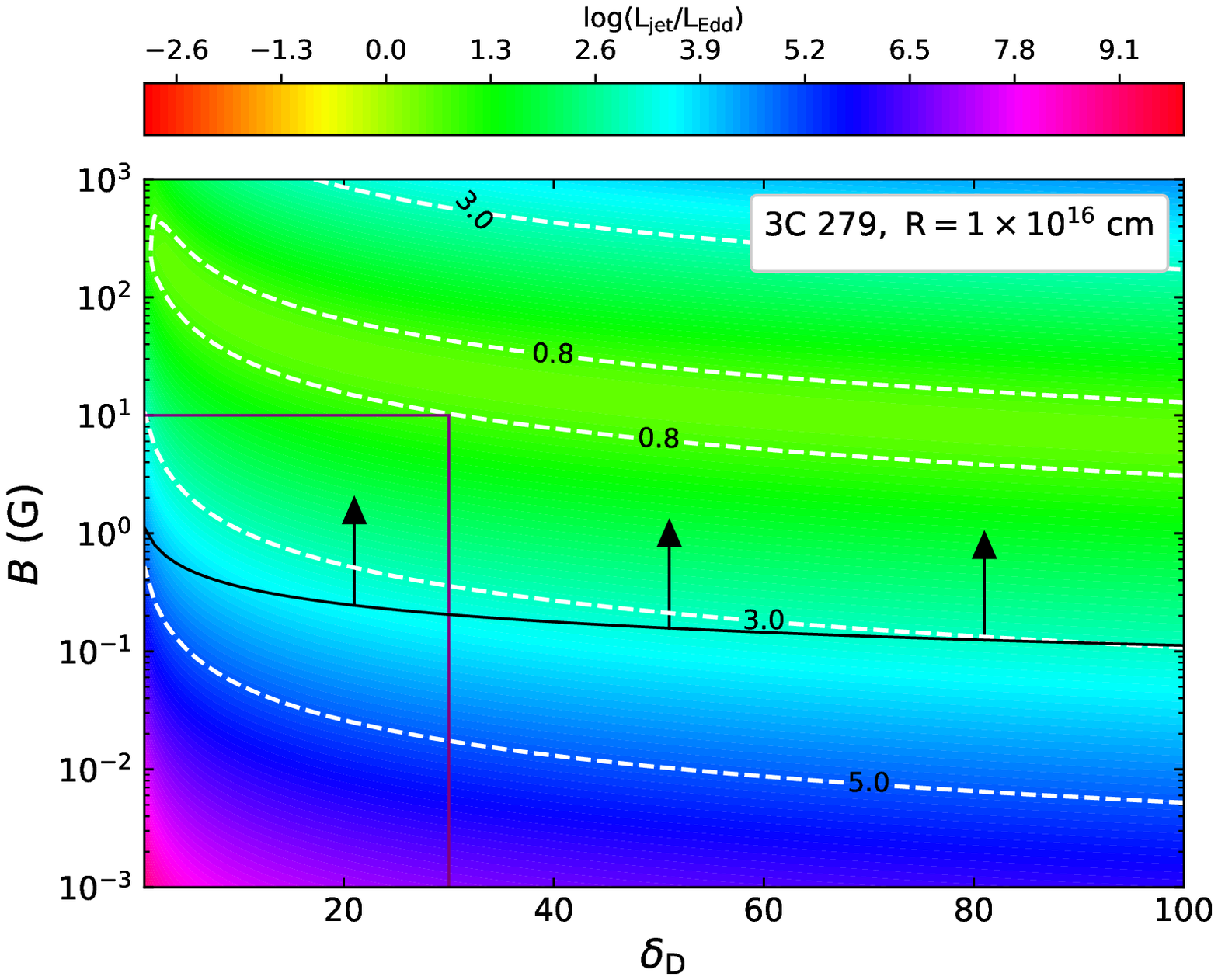}
}\hspace{-5mm}
\vspace{-1mm}
\quad
\subfloat{
\includegraphics[width=0.99\columnwidth]{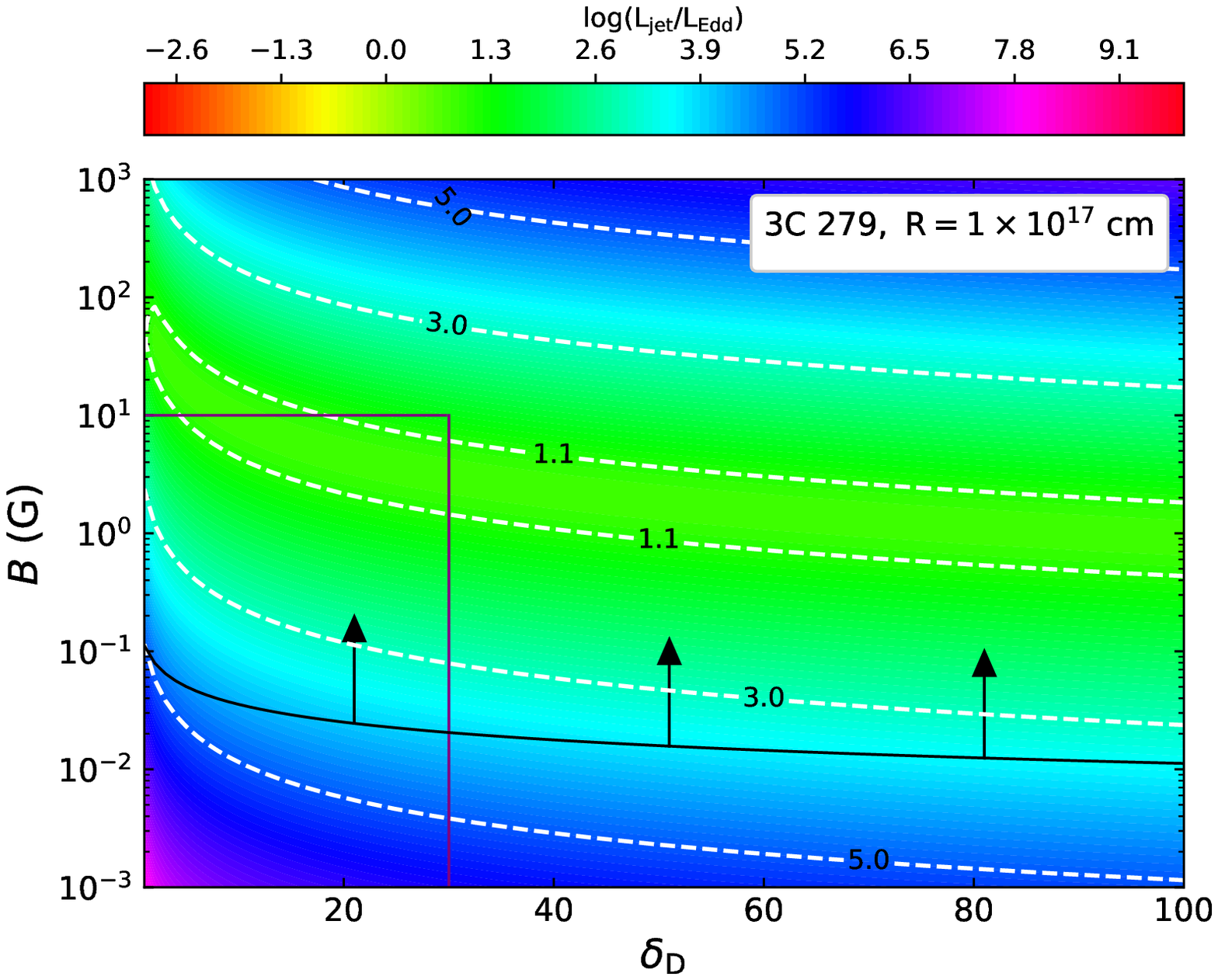}
}
\caption{The ratio of $L_{\rm jet}/L_{\rm Edd}$ in the $\delta_{\rm D}$-$B$ diagram for 3C 279 peaking at 0.2~GeV, when setting $R=1\times 10^{14},~1\times 10^{15},~1\times 10^{16},~1\times 10^{17}~\rm cm$, respectively. The line styles in all panels have the same meaning as in Fig.~\ref{0229}. \label{279-02G}}
\end{figure*}

\begin{figure*}
\centering
\subfloat{
\includegraphics[width=1\columnwidth]{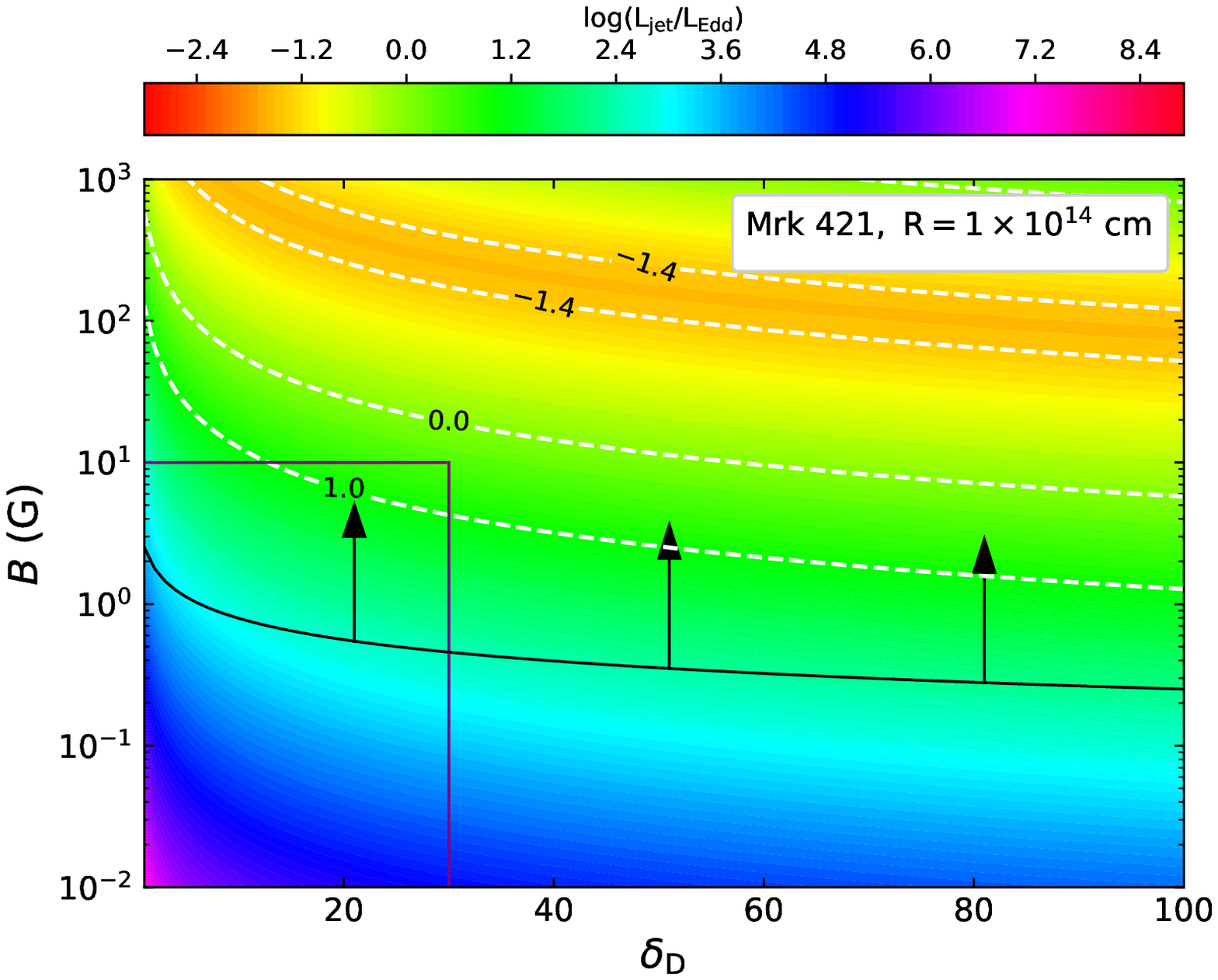}
}\hspace{-5mm}
\vspace{-1mm}
\quad
\subfloat{
\includegraphics[width=1\columnwidth]{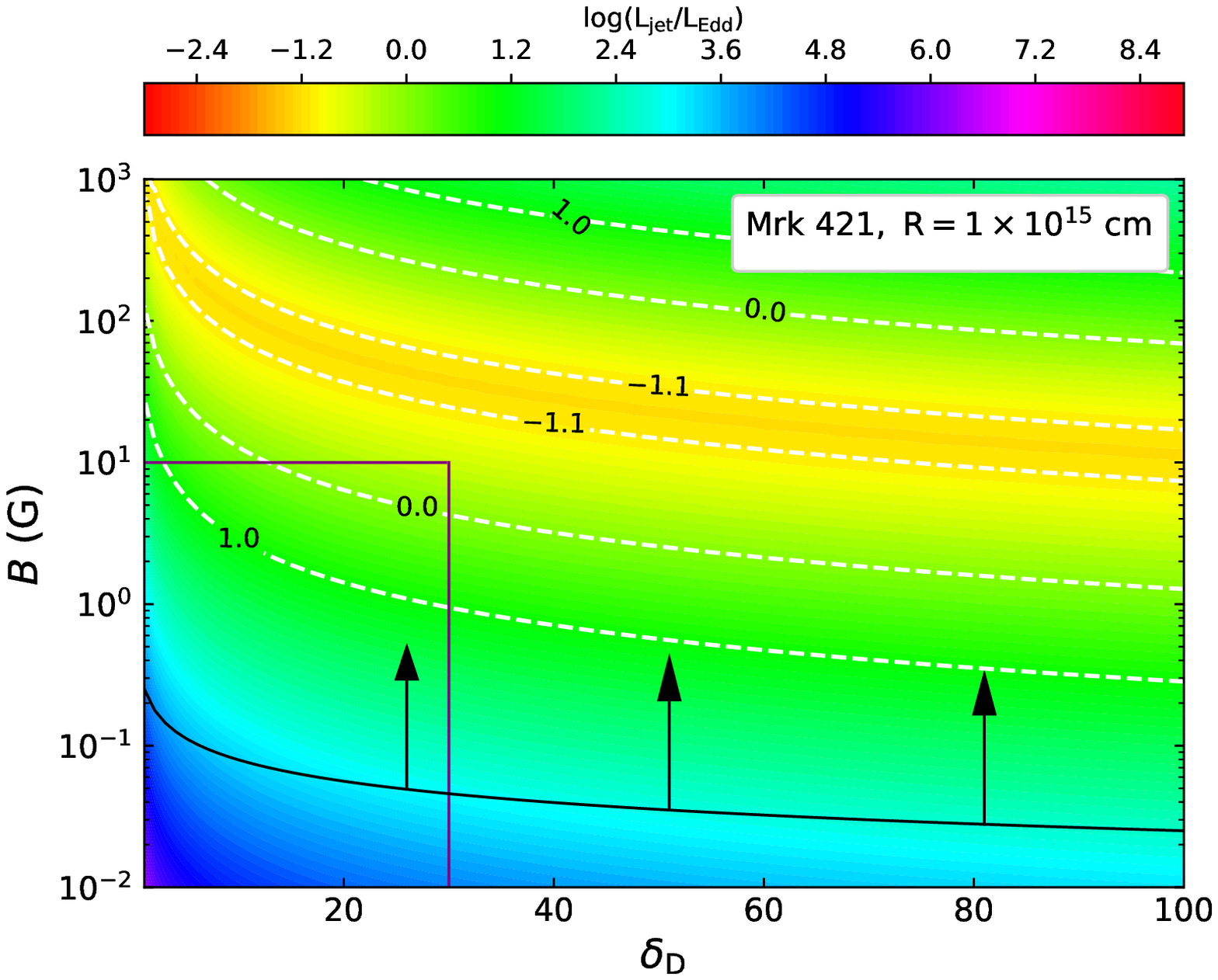}
}\hspace{-5mm}
\vspace{-1mm}
\quad
\subfloat{
\includegraphics[width=0.99\columnwidth]{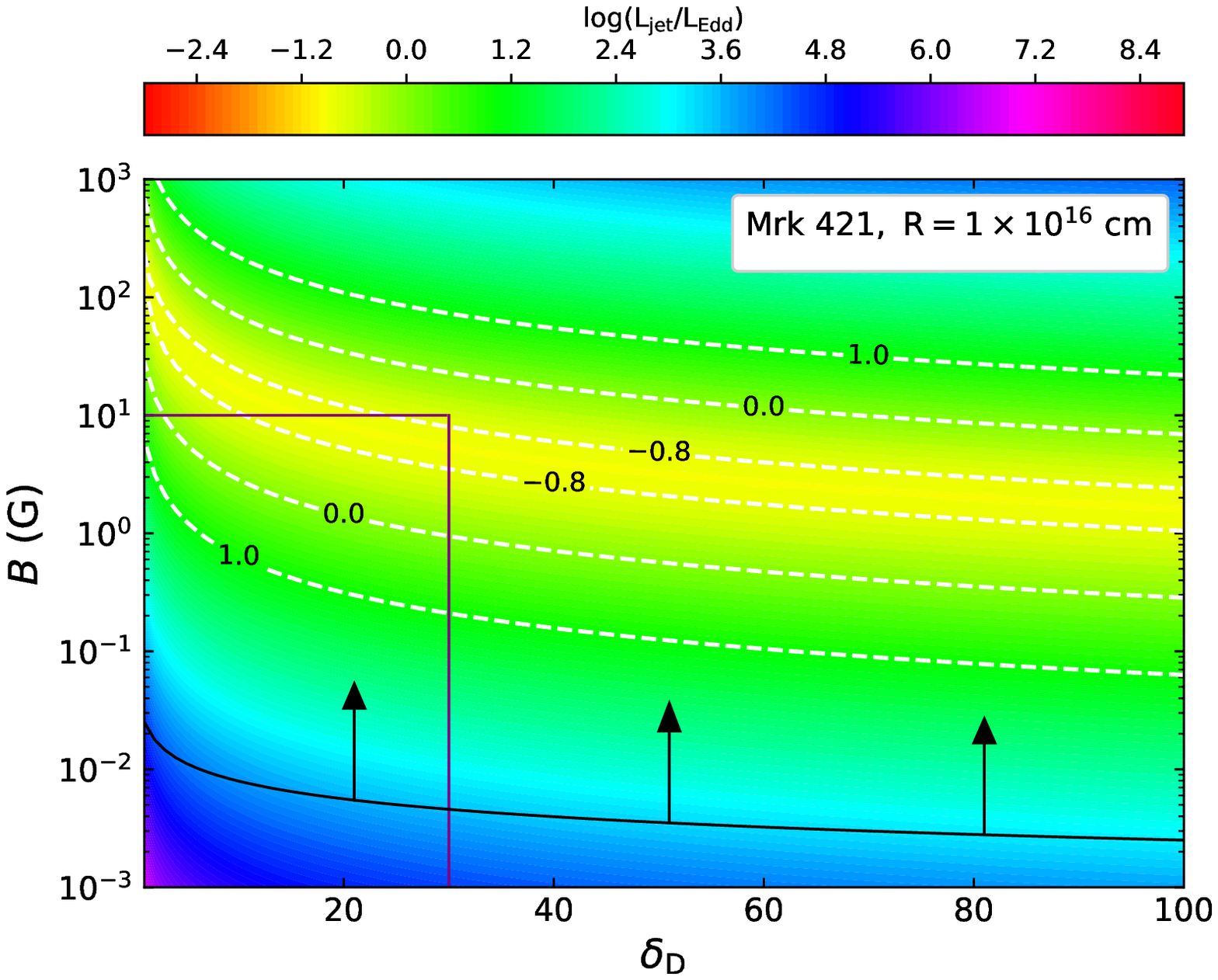}
}\hspace{-5mm}
\vspace{-1mm}
\quad
\subfloat{
\includegraphics[width=0.99\columnwidth]{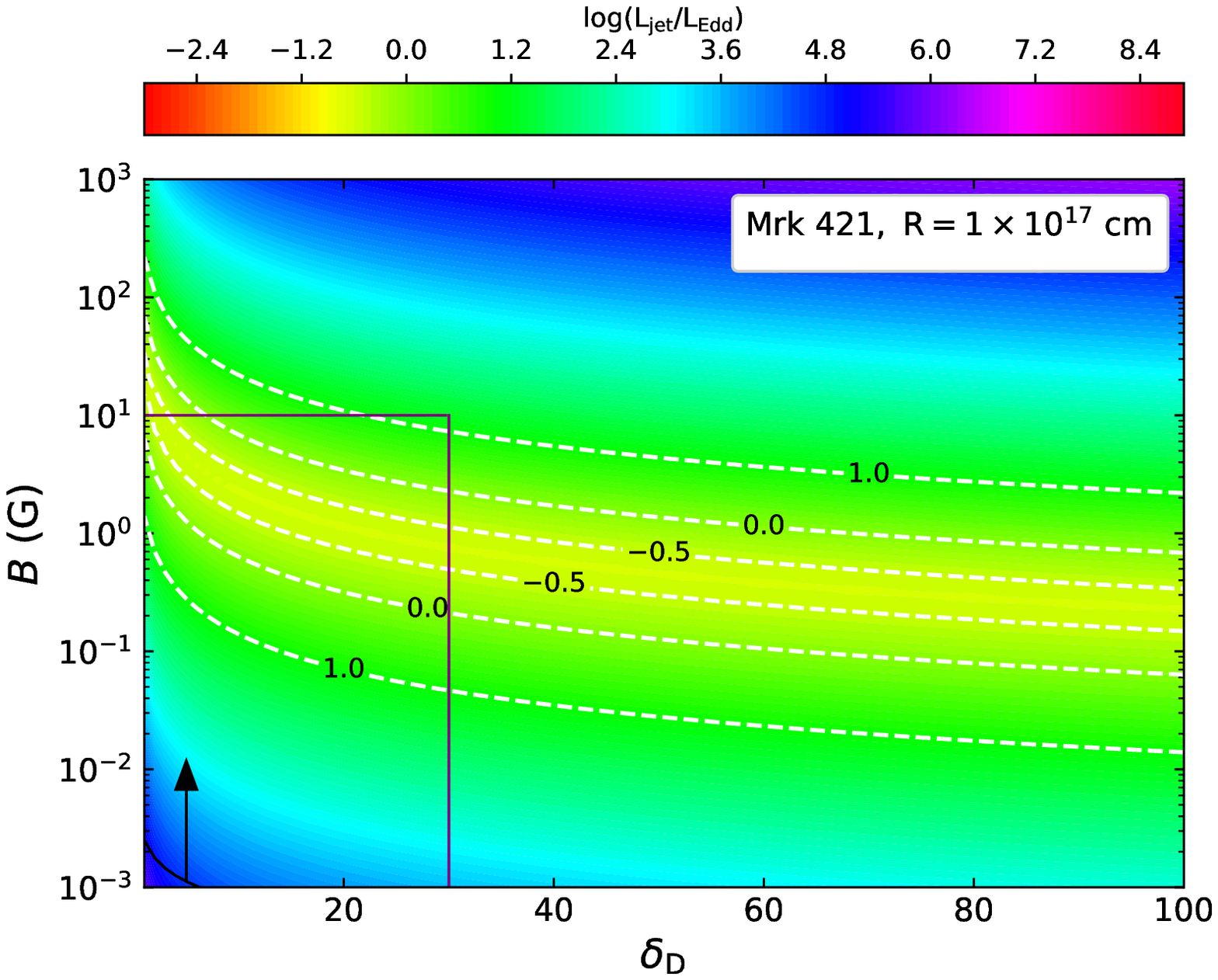}
}\hspace{-5mm}
\vspace{-1mm}
\quad
\subfloat{
\includegraphics[width=0.99\columnwidth]{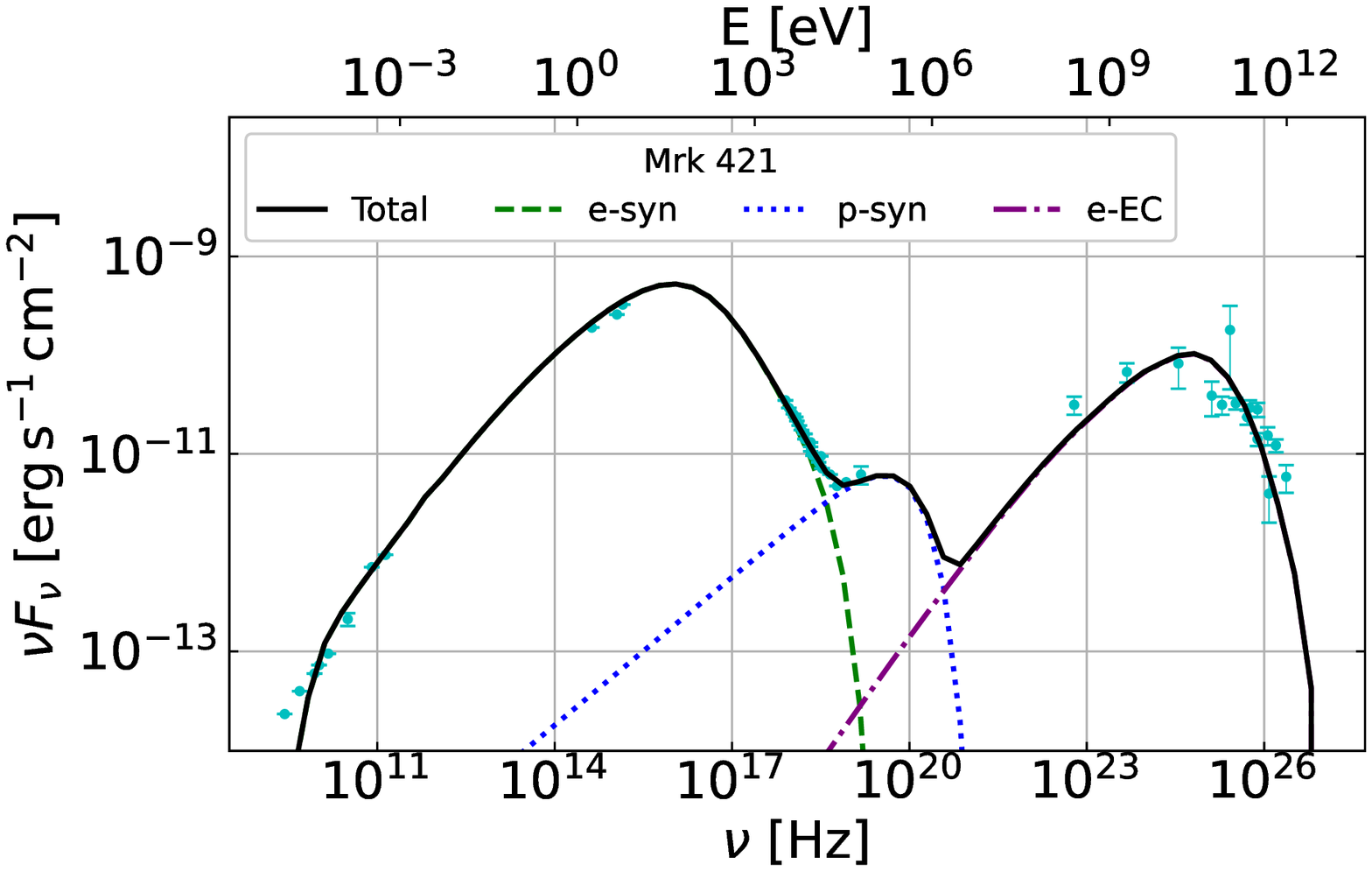}
}\hspace{-5mm}
\vspace{-1mm}
\quad
\subfloat{
\includegraphics[width=0.99\columnwidth]{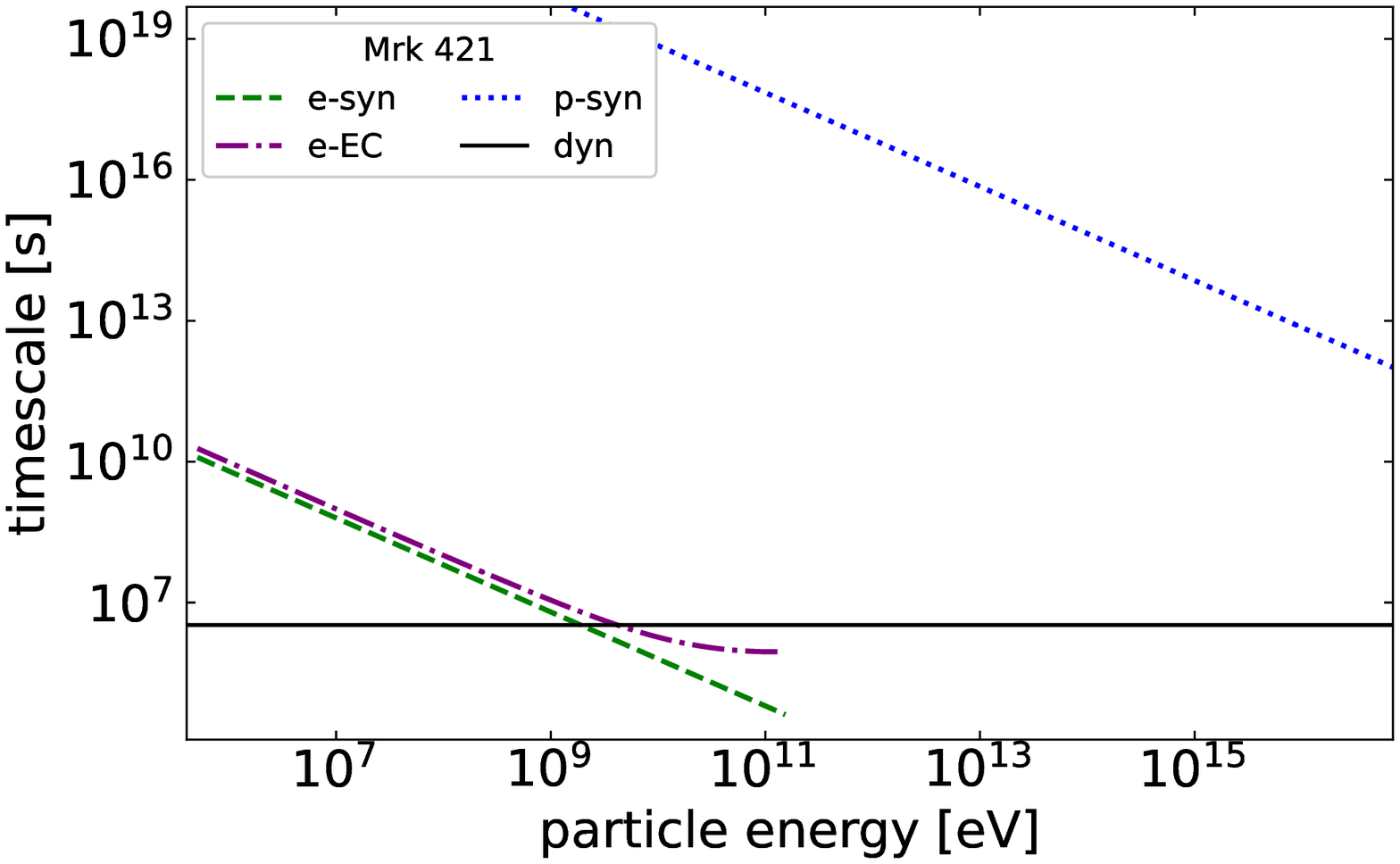}
}
\caption{Upper four panels: The ratio of $L_{\rm jet}/L_{\rm Edd}$ in the $\delta_{\rm D}$-$B$ diagram for Mrk 421 peaking at 100~keV, when setting $R=1\times 10^{14},~1\times 10^{15},~1\times 10^{16},~1\times 10^{17}~\rm cm$, respectively. Lower two panels: The fitting result of the quasi-simultaneous SED of Mrk 421 with the conventional one-zone model (left panel) and the corresponding timescales of various cooling processes for relativistic electrons and protons as a function of particle energy in the comoving frame (right panel). In the lower left panel, the cyan points are data taken from \cite{2016ApJ...827...55K}. The parameters are the same as those shown in Table~\ref{parameters}. In the lower right panel, purple dash-dotted line represent the relativistic electrons EC emission. Other line styles in all panels have the same meaning as in Fig.~\ref{0229}. \label{421-100k}}
\end{figure*}

\begin{figure*}
\centering
\subfloat{
\includegraphics[width=1\columnwidth]{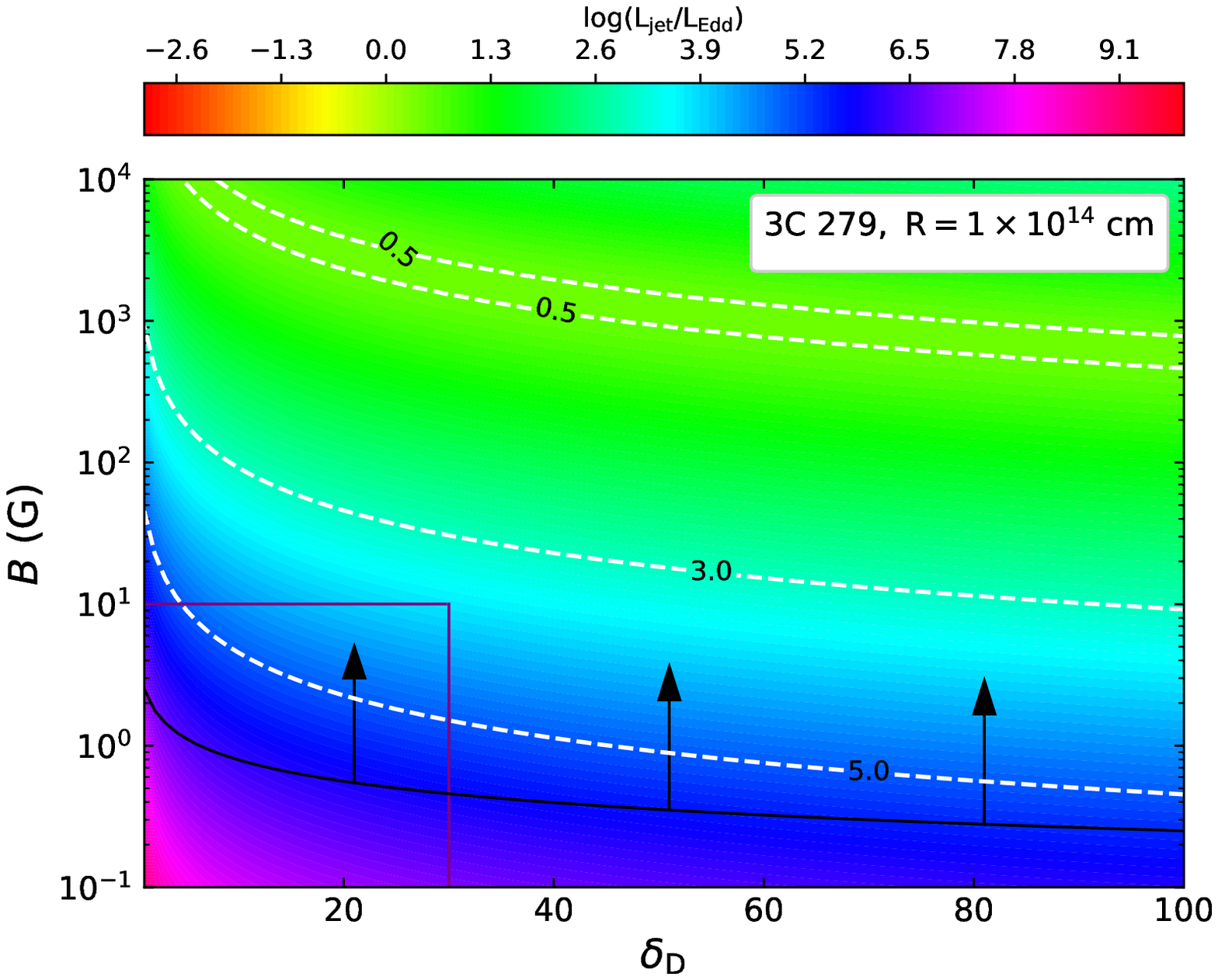}
}\hspace{-5mm}
\vspace{-1mm}
\quad
\subfloat{
\includegraphics[width=1\columnwidth]{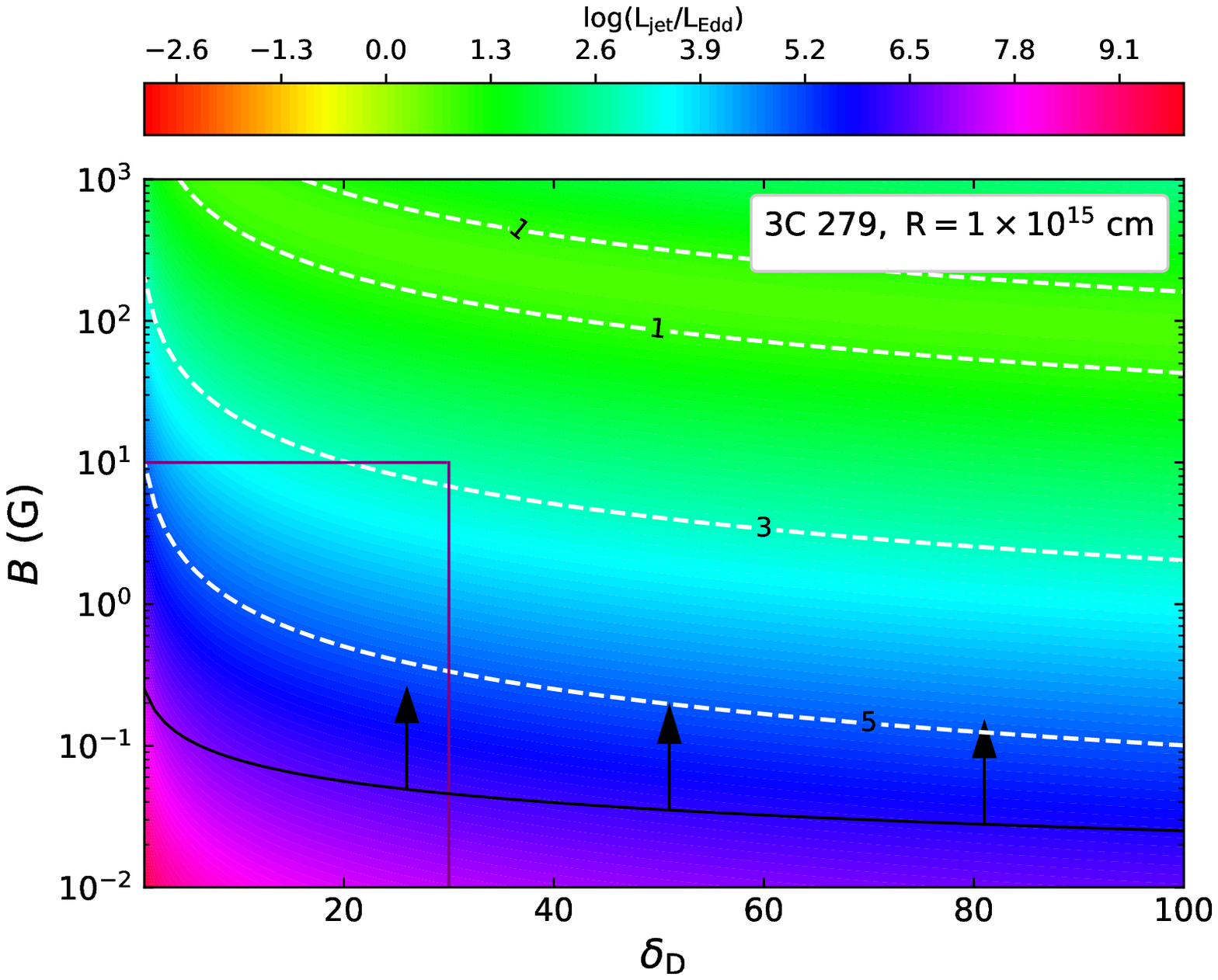}
}\hspace{-5mm}
\vspace{-1mm}
\quad
\subfloat{
\includegraphics[width=0.99\columnwidth]{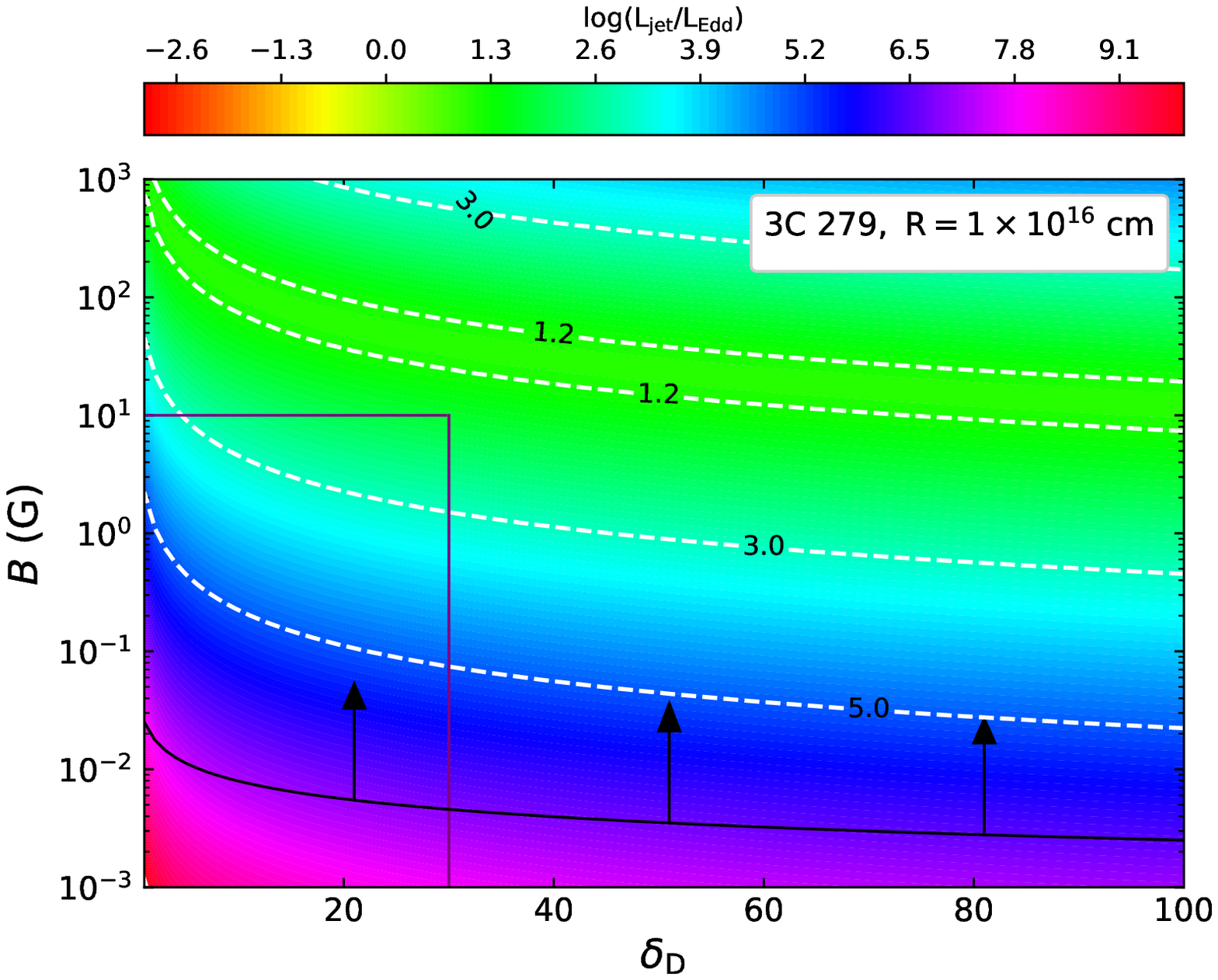}
}\hspace{-5mm}
\vspace{-1mm}
\quad
\subfloat{
\includegraphics[width=0.99\columnwidth]{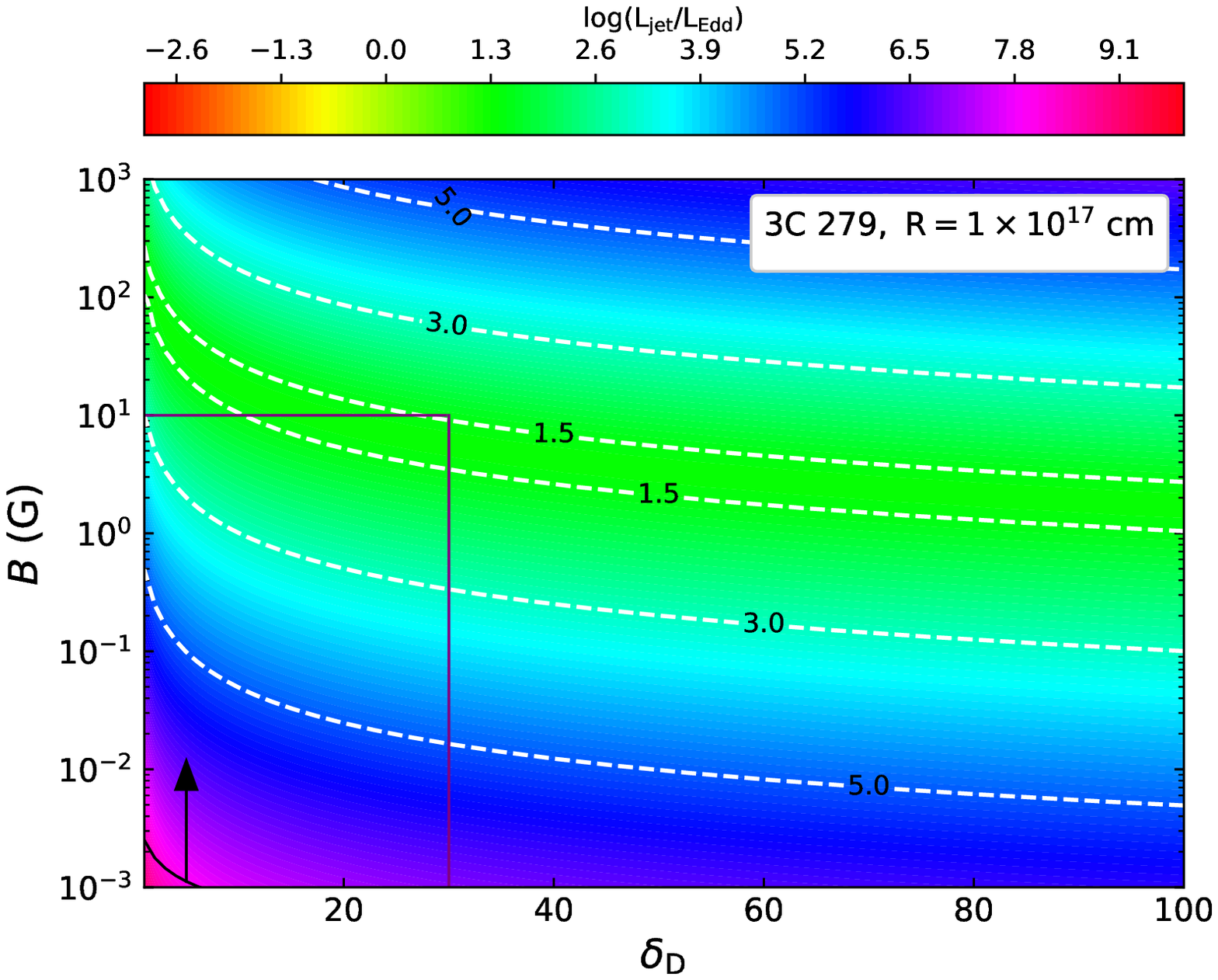}
}
\caption{The ratio of $L_{\rm jet}/L_{\rm Edd}$ in the $\delta_{\rm D}$-$B$ diagram for 3C 279 peaking at 100~keV, when setting $R=1\times 10^{14},~1\times 10^{15},~1\times 10^{16},~1\times 10^{17}~\rm cm$, respectively. The line styles in all panels have the same meaning as in Fig.~\ref{0229}. \label{279-100k}}
\end{figure*}

In the following, we study the parameter space when assuming proton synchrotron radiation dominants the high-energy hump that peaks in four energy ranges, which are 10~keV$\sim$100~keV, 0.1~GeV$\sim$10~GeV, 10~GeV$\sim$1~TeV, and 1~TeV$\sim$10~TeV, respectively. Note that, these peak energies are given after correcting for the extragalactic background light (EBL) absorption \cite{2011MNRAS.410.2556D}. In each energy range, two blazars are studied as representatives. The detailed information of the sample is given in Table~\ref{sample}. The derived parameter spaces, corresponding fitting results and cooling timescales are shown in Figs.~\ref{0229}--\ref{279-100k}. 

\textit{Case 1: 1~TeV$\sim$10~TeV.} Here we study two typical hard-TeV spectrum blazars, which are 1ES 0229+200 and 1ES 1101-232. In the $\delta_{\rm D}$--$B$ diagrams of Figs.~\ref{0229} and \ref{1101}, under the observational constraints, it can be seen that only when setting $R=1\times 10^{17}~\rm cm$ the parameter space can be found. Especially for 1ES 0229+200, its parameter space is restricted strictly. In the modeling, the high-energy hump is dominated by the proton synchrotron radiation, and the SSC radiation of electrons can be ignored, since electrons mainly lose energy through the synchrotron radiation, as shown by the cooling timescales in Figs.~\ref{0229} and \ref{1101}. Please note that, the UV data points of 1ES 0229+200 are suggested as the emission from the host galaxy \cite{2018MNRAS.477.4257C}, therefore we do not fit them with the jet model.

\textit{Case 2: 10~GeV$\sim$1~TeV.} In this energy range, we study 1ES 2037+521 and Mrk 421. The derived results are basically similar to those obtained in the range of 1~TeV$\sim$10~TeV. As shown in Table~\ref{sample}, since the high-energy hump's peak luminosities of 1ES 2037+521 and Mrk 421 are lower than those of 1ES 0229+200 and 1ES 1101-232, larger parameter spaces are derived. It can be seen in Figs.~\ref{2037} and \ref{421-0.1T} that parameter spaces can be found under the observational constraints when $R=1\times 10^{16}~\rm cm$ and $R=1\times 10^{17}~\rm cm$. In the modeling, their high-energy humps are fitted by the proton synchrotron radiation as well.

\textit{Case 3: 0.1~GeV$\sim$10~GeV.} Here we study two powerful blazars, which are S5 0716+714 and 3C 279. Compared to the blazars studied in $Case~1$ and $Case~2$, they have higher peak luminosities and lower peak energies. It can be found in Figs.~\ref{S5} and \ref{279-02G} that the super-Eddington jet power is generally needed if applying the proton synchrotron radiation to explain their high-energy humps. Only when setting $R=1\times10^{14}\rm~cm$, 3C 279 can have a small parameter space. However if considering $\delta_{\rm D}\lesssim30$, it is necessary to introduce an extreme magnetic field exceeding $10^3$~G, which is obviously contrary to the observation. Therefore, we do not fit their SEDs with the one-zone proton synchrotron model due to the requirement of extreme physical parameters. A reasonable interpretation can be given by the conventional leptonic model \cite{2013ApJ...768...54B}. Blazars with high-energy hump peaks in the range of 0.1~GeV$\sim$10~GeV are usually low or intermediate synchrotron peaked blazars \cite{2010ApJ...716...30A, 2016ApJS..226...20F, 2022ApJS..262...18Y}. Following the $Fermi$ blazar sequence \cite{2017MNRAS.469..255G}, low or intermediate synchrotron peaked blazars usually have high $\gamma$-ray luminosities. Compared to $Case~1$ and $Case~2$, the radiation in the range of 0.1~GeV$\sim$10~GeV is produced by lower energy protons with lower radiation efficiency, therefore a higher proton injection luminosity, exceeding the Eddington luminosity, is required.

\textit{Case 4: 10~keV$\sim$100~keV.} In this energy range, we study if the hard X-ray components of Mrk 421 and 3C 279 can be explained by the proton synchrotron radiation. As discussed in $Case~3$, hard X-ray spectra are more difficult to interpret with proton synchrotron radiation because it is contributed by lower energy protons which are cooled inefficient. For the hard X-ray component of Mrk 421, \cite{2017ApJ...842..129C} argues that it cannot be interpreted by the low-energy tail of the SSC emission in the framework of one-zone SSC model. Here, we study if the proton synchrotron radiation could be a possible explanation. In Fig.~\ref{421-100k}, it can be seen that, since the hard X-ray luminosity of Mrk 421 is quite low, parameter space under the observational constraints can be found. By employing the values of $\delta_{\rm D}$, $B$, and $R$ in the obtained parameter space, we fit the hard X-ray spectrum of Mrk 421 with the proton synchrotron radiation (see \cite{2017ApJ...842..129C, 2022A&A...668A...3B, 2022PhRvD.106j3021X} for other explanations). However, with the adopted blob radius, the energy density of synchrotron photons emitted by relativistic electrons $U_{\rm e, syn}\approx 3.2\times10^{-6}~\rm erg~cm^{-3}$ would be much lower than that of magnetic field $U_{\rm B}\approx 2.5\times10^{-3}~\rm erg~cm^{-3}$. Therefore, in the framework of one-zone model, additional radiation component should be introduced to explain the high-energy hump instead of the SSC radiation. We find that the broad-line region luminosity $L_{\rm BLR}\approx 2.3\times10^{41}~\rm erg~s^{-1}$ of Mrk 421 is given in \cite{2012ApJ...759..114C}, which may suggest the existence of weak external photon fields. In this work, we assume that the blob is placed at $\sim 0.1~\rm pc$ away from the SMBH, so that the EC radiation can fit the high-energy hump (here the energy densities of external photon fields are calculated with Eq. A4 and A5 in \cite{2022PhRvD.106j3021X}). For 3C 279, because of its high hard X-ray luminosity, no parameter space is found as shown in Fig.~\ref{279-100k}.

\begin{table*}
\setlength\tabcolsep{11pt}
\caption{The Sample. Columns from left to right: (1) the energy range that the high-energy hump peaks in. (2) the object name. (3) the redshift of the object. (4) the SMBH mass in units of the solar mass, $M_{\odot}$. (5) the object's EBL corrected peak energy of high-energy hump. (6) the object's EBL corrected peak luminosity of high-energy hump. (7) references that provide $E_{\rm peak}^{\rm obs}$ and ${\nu L_\nu}_{\rm peak}^{\rm obs}$. For 1ES 1101-232 and 1ES 2037+521, in absence of an estimated black hole mass, we considered an average value of $10^9 M_\odot$ \citep{2017ApJ...851...33P, 2022ApJ...925...40X}.}\label{sample}
\centering
\begin{tabular}{ccccccc}
\hline\hline														
Energy range	&	Object	&	$z$	&	$M_{\rm BH}$	&	$E_{\rm peak}^{\rm obs}$	&	${\nu L_\nu}_{\rm peak}^{\rm obs}$	&	Refs	\\	
	(1) & (2) & (3) & (4) & (5) &(6)& (7)  \\
\hline														
1~TeV$\sim$10~TeV	&	1ES 0229+200	&	0.14	&	$1.5\times10^{9}$ \citep{2012AA...542A..59M}	&	7.3~TeV	&	$1.1\times10^{45}~\rm erg~s^{-1}$	&	\cite{2014ApJ...782...13A}	\\	
	&	1ES 1101-232	&	0.188	&	9	&	1.1~TeV	&	$1.0\times10^{45}~\rm erg~s^{-1}$	&	\cite{2007AA...470..475A}	\\	
\hline														
10~GeV$\sim$1~TeV	&	1ES 2037+521	&	0.053	&	9	&	140~GeV	&	$1.9\times10^{43}~\rm erg~s^{-1}$	&	\cite{2020ApJS..247...16A}	\\	
	&	Mrk 421	&	0.031	&	$1.3\times10^9$ \citep{2002AA...389..742W}	&	50~GeV	&	$1.8\times10^{44}~\rm erg~s^{-1}$	&	\cite{2016ApJ...827...55K}	\\	
\hline														
0.1~GeV$\sim$10~GeV	&	S5 0716+714	&	0.31	&	$1\times10^8$ \citep{2015MNRAS.450L..21Z}	&	1~GeV	&	$1.4\times10^{46}~\rm erg~s^{-1}$	&	\cite{2013ApJ...768...54B}	\\	
	&	3C 279	&	0.536	&	$8\times10^8$ \citep{2015MNRAS.450L..21Z}	&	0.2~GeV	&	$7.2\times10^{46}~\rm erg~s^{-1}$	&	\cite{2013ApJ...768...54B}	\\	
\hline														
10~keV$\sim$100~keV	&	Mrk 421	&	0.031	&	$1.3\times10^9$ \citep{2002AA...389..742W}	&	100~keV	&	$1.4\times10^{43}~\rm erg~s^{-1}$	&	\cite{2016ApJ...827...55K}	\\	
	&	3C 279	&	0.536	&	$8\times10^8$ \citep{2015MNRAS.450L..21Z}	&	100~keV	&	$1.2\times10^{46}~\rm erg~s^{-1}$	&	\cite{2013ApJ...768...54B}	\\	
\hline														
\label{sample}
\end{tabular}
\end{table*}

\begin{table*}
\setlength\tabcolsep{7pt}
\caption{Parameters for the SED fitting with the conventional one-zone model. Columns from left to right: (1) the object name. (2) the Doppler factor. (3) the magnetic field. (4) the blob radius. (5) the injection luminosity of relativistic electrons in the comoving frame. (6) the break electron Lorentz factor. (7) the electron spectral index before $\gamma_{\rm e,b}$. (8) the electron spectral index after $\gamma_{\rm e,b}$. (9) the injection luminosity of relativistic protons in the comoving frame. (10) the maximum proton Lorentz factor. Note that, in the modeling of leptonic emission, the minimum and maximum electron Lorentz factors are fixed to 1 and $10^6$, since they have little effect on the fitting results. For relativistic protons, the proton minimum Lorentz factor is fixed to 1. The adopted value of maximum proton Lorentz factor $\gamma_{\rm p,max}$ is lower than that constrained by the Hillas condition. The magnetic field $B$ and Doppler factor $\delta_{\rm D}$ are selected from the parameter space that obtained in corresponding $\delta_{\rm D}$--$B$ diagrams.}
\centering
\begin{tabular}{cccccccccc}
\hline\hline																			
Object 	&	$\delta_{\rm D}$	&	$B$	&	$R$	&	$L_{\rm e,inj}$	&	$\gamma_{\rm e,b}$	&	$\alpha_{\rm e,1}$	&	$\alpha_{\rm e,2}$	&	$L_{\rm p,inj}$	&	$\gamma_{\rm p,max}$	\\
	&		&	(G)	&	(cm)	&	$(\rm erg~s^{-1})$	&		&		&		&	$(\rm erg~s^{-1})$	&		\\
	(1) & (2) & (3) & (4) & (5) &(6)& (7) & (8) & (9) & (10) \\
\hline																			
1ES 0229+200	&	7	&	10	&	$1\times10^{17}$	&	$1.7\times10^{42}$	&	$1.8\times10^{5}$	&	1.1	&	4.1	&	$5.7\times10^{42}$	&	$3.4\times10^{11}$	\\
1ES 1101-232	&	10	&	3	&	$1\times10^{17}$	&	$1.3\times10^{42}$	&	$7.5\times10^{4}$	&	1.1	&	3.7	&	$1.3\times10^{43}$	&	$1.0\times10^{11}$	\\
1ES 2037+521	&	20	&	2	&	$1\times10^{17}$	&	$2.2\times10^{39}$	&	$1.5\times10^{5}$	&	1.1	&	3.8	&	$7.5\times10^{40}$	&	$6.9\times10^{10}$	\\
Mrk 421 ($E_{\rm peak}^{\rm obs}=50~\rm GeV$)	&	20	&	1	&	$1\times10^{17}$	&	$2.9\times10^{40}$	&	$1.5\times10^{4}$	&	1.2	&	3.8	&	$3.8\times10^{43}$	&	$2.8\times10^{10}$	\\
Mrk 421 ($E_{\rm peak}^{\rm obs}=100~\rm keV$)	&	26	&	0.25	&	$1\times10^{17}$	&	$2.4\times10^{40}$	&	$3.1\times10^{4}$	&	1.3	&	4.3	&	$2.7\times10^{44}$	&	$8.6\times10^{7}$	\\
\hline
\label{parameters}
\end{tabular}
\end{table*}

\section{Summary}\label{sum}
In this work, we revisit the proton synchrotron radiation in blazar jets and study its possible contribution on the high-energy hump. The analytical analysis is based on the following three constraints.
\begin{enumerate}
\item The total jet power does not exceed the Eddington luminosity of SMBH.
\item The maximum proton energy is smaller than that obtained from the Hillas condition.
\item The emitting region is transparent to the $\gamma$-ray emission. 
\end{enumerate}
We present the ratio of $L_{\rm jet}/L_{\rm Edd}$ in the $B$--$\delta_{\rm D}$ parameter space for different blob radius. As studied in $Case~1$ and $Case~2$, the parameter space can be found for $>10$ GeV emission. On the premise that the jet power does not exceed the Eddington luminosity, if considering a relative large blob radius ($R=1\times10^{16}\rm~or~1\times10^{17}~\rm cm$), the parameter space that satisfies observational constraints (i.e., $B\lesssim10~\rm G$ and $\delta_{\rm D}\lesssim30$) can be found, and if assuming a relative compact blob ($R=1\times10^{14} \rm~or~1\times10^{15}~\rm cm$), a strong magnetic field $>100~\rm G$ is inevitably needed. This result is consistent with \cite{2016ApJ...825L..11P}. Therefore, by considering a canonical value of the Doppler factor, i.e., $\delta_{\rm D}=10$, we suggest that proton synchrotron radiation can account for the variability of $>10$ GeV emission longer than day-scale. On the other hand, when interpreting the faster variability, the proton synchrotron radiation model can be ruled out (also see the discussion in \cite{2015MNRAS.448..910C}) if it is agreed that extremely strong magnetic fields do not exist in blazar jet. For the powerful $\gamma$-ray emission peaked in the range of 0.1 GeV -- 10 GeV studied in $Case~3$, no parameter space is found, as concluded in \cite{2015MNRAS.450L..21Z, 2016ApJ...825L..11P}. We want to emphasize that if the GeV luminosities of some blazars are weak enough, proton synchrotron radiation may still be able to explain the high-energy hump. For the 10 keV -- 100 keV emission studied in $Case~4$, the parameter space that satisfies the observational constraints can be found if the object's luminosity is low enough, however it still cannot to find a parameter space when the object has a powerful keV emission.

To summarize, our results suggest that, under certain conditions, proton synchrotron radiation is a possible explanation for the high-energy emission of blazars. On the other hand, since proton synchrotron radiation predicts a higher maximum degree of polarization than that of inverse Compton scattering \cite{2013ApJ...774...18Z, 2018ApJ...863...98P, 2019ApJ...876..109Z}, recent and future X-ray and $\gamma$-ray polarization observations \cite{2022icrc.confE.580D, 2022ApJ...938L...7D, 2022Natur.611..677L, 2022JATIS...8b6002W} will be important to distinguish the leptonic and proton synchrotron emission at high-energy bands.

\section*{Acknowledgements}
We thank the anonymous referees for insightful comments and constructive suggestions. We sincerely thank Dr. Luigi Costamante for sharing the SED data. This work is supported by the National Natural Science Foundation of China (NSFC) under the grants No. 12203043. H.B.X. acknowledges the support by the NSFC under the the Grants No. 12203034 and by Shanghai Science and Technology Fund under the Grants No. 22YF1431500. Z.R.W. acknowledges the support by the NSFC under the the Grants No. 12203024.

\bibliographystyle{apsrev}
\bibliography{ms.bib}

\end{document}